\documentclass[12pt]{article}
\usepackage{lscape}
\usepackage{geometry}
\usepackage[onehalfspacing]{setspace}
\usepackage{amssymb}
\usepackage{amsfonts}
\usepackage{graphicx}
\usepackage{amsmath}
\usepackage{epstopdf}
\usepackage{natbib}
\usepackage{morefloats}
\usepackage{float}
\usepackage{comment}
\usepackage{pdflscape}
\usepackage{adjustbox}
\usepackage{subcaption}
\usepackage{longtable}
\usepackage{apalike}
\usepackage{xfrac}
\usepackage[dvipsnames,table]{xcolor}
\usepackage{etoolbox,siunitx}
\robustify\bfseries
\usepackage{xstring}

\usepackage{qtree}
\usepackage{multicol}
\usepackage{booktabs}
\usepackage{threeparttable}
\usepackage{tabularx}
\usepackage{dcolumn}
\newcolumntype{d}[1]{D..{#1}} 
 
\usepackage{siunitx}
 \usepackage{mathpazo} 
\usepackage[scaled]{helvet} 
\usepackage{courier} 
\normalfont
\usepackage[T1]{fontenc}
\usepackage{listings}
\usepackage{epigraph}
\def\sym#1{\ifmmode^{#1}\else\(^{#1}\)\fi}

\usepackage{xcolor} 
\definecolor{DarkerPineGreen}{RGB}{0, 90, 80} 
\definecolor{dukeblue}{rgb}{0.0, 0.0, 0.61}
\definecolor{darkblue}{RGB}{10, 10, 100}
 \definecolor{siena}{rgb}{0.91,0.45,0.32} 
\definecolor{darkred}{rgb}{0.8,0,0}
\definecolor{darkpowderblue}{rgb}{0.0, 0.05, 0.5}
 \definecolor{dpd2}{rgb}{0.0, 0.043, 0.43}
 \definecolor{dpd}{rgb}{0.0, 0.05, 0.5}
 
\definecolor{darkblue2}{HTML}{1e3986}
\definecolor{darkred2}{HTML}{B22222}
\definecolor{darkgreen2}{HTML}{1B9E77}

\usepackage[pdftex,bookmarks,colorlinks]{hyperref}
\hypersetup{
  colorlinks,
  citecolor=darkpowderblue,
  linkcolor=red,
  urlcolor=blue}
\usepackage{cleveref}

\usepackage{etoolbox}
 
\makeatletter
\patchcmd{\epigraph}{\@epitext{#1}}{\itshape\@epitext{#1}}{}{}
\makeatother

\makeatletter
\newcommand\halftiny{\@setfontsize\halftiny\@vipt\@viipt}
\newcommand\notsotiny{\@setfontsize\notsotiny{6.99}{9.2828}}
\newcommand\footscript{\@setfontsize\footscript{8.5}{10}}
\newcommand\footscriptfuck{\@setfontsize\footscriptfuck{9.5}{11}}
\newcommand\notsolarge{\@setfontsize\notsolarge{12}{14}}
\newcommand\superlarge{\@setfontsize\superlarge{21.9}{24}}
\makeatother
 
\renewenvironment{abstract}
 {\small
  \begin{center}
  \bfseries \abstractname\vspace{-.5em}\vspace{0pt}
  \end{center}
  \list{}{
    \setlength{\leftmargin}{2cm}    \setlength{\rightmargin}{\leftmargin}  }  \item\relax}
 {\endlist}
\usepackage{graphicx}
\usepackage{wrapfig}
 
\allowdisplaybreaks
 
\usepackage{enumitem}

\usepackage{algorithm,algcompatible}
\usepackage{tikz}
\usetikzlibrary{positioning}

\begin{document}
\sloppy

    \title{\vspace*{-0.25cm} \superlarge \textbf{\color{dpd} Dual Interpretation of Machine Learning Forecasts} \vspace*{-0.5cm}}
\author{\hspace*{1.7cm} Philippe Goulet Coulombe\thanks{  \footscriptfuck 
Contact: \href{mailto:p.gouletcoulombe@gmail.com}{\texttt{\color{black} goulet\_coulombe.philippe@uqam.ca}}. For helpful comments,  we thank Mikael Frenette, Nicolas Harvie, Julien Martin, Francesco Simone Lucidi, Tim Reinicke, Gabriel Rodriguez Rondon, Dalibor Stevanovic, Anne Valder, and Boyuan Zhang. The views expressed in this paper do not necessarily reflect those of the Oesterreichische Nationalbank or the Eurosystem.  Goulet Coulombe gratefully acknowledges funding from the OeNB's Klaus Liebscher Economic Research Scholarship for this work.  R codes are available on \href{https://github.com/maxi-tb22/DualML}{Github}. This draft: \today. }\\[-0.2cm] 
  \hspace*{1.7cm} \textbf{\texttt{\fontfamily{phv}\selectfont \notsolarge Université du Québec à Montréal}} 
  \and 
  \\[1.3cm] \hspace*{-9.6cm} Maximilian G\"obel\\[-0.2cm] \hspace*{-9.6cm} \textbf{\texttt{\fontfamily{phv}\selectfont \footnotesize\notsolarge Brain}} 
  \\[0.5cm] \hspace*{-9.6cm} Karin Klieber\\[-0.2cm] \hspace*{-9.6cm} \textbf{\texttt{\fontfamily{phv}\selectfont \footnotesize\notsolarge Oesterreichische Nationalbank}}
}
\date{\vspace{0.2cm}
\small
\small
\vspace{-0.4cm}
\large
  }
  \newgeometry{left=1.9 cm, right= 1.9 cm, top=2.3 cm, bottom=1.5 cm}
\maketitle
\begin{abstract}

\noindent Machine learning predictions are typically interpreted as the sum of contributions of predictors. Yet, each out-of-sample prediction can also be expressed as a linear combination of in-sample values of the predicted variable, with weights corresponding to pairwise proximity scores between current and past economic events. While this dual route leads nowhere in some contexts (e.g., large cross-sectional datasets), it provides sparser interpretations in settings with many regressors and little training data---like macroeconomic forecasting. In this case, the sequence of contributions can be visualized as a time series, allowing analysts to explain predictions as quantifiable combinations of historical analogies. Moreover, the weights can be viewed as those of a data portfolio, inspiring new diagnostic measures such as forecast concentration, short position, and turnover. We show how weights can be retrieved seamlessly for (kernel) ridge regression, random forest, boosted trees, and neural networks.  Then, we apply these tools to analyze post-pandemic forecasts of inflation, GDP growth, and recession probabilities. In all cases, the approach opens the black box from a new angle and demonstrates how machine learning models leverage history partly repeating itself. \\
\phantom{ouhouh}

\end{abstract}
 
\thispagestyle{empty}
 
 
 
 
 
\clearpage
 
 
\clearpage 
\setcounter{page}{1}
\restoregeometry

\section{Introduction}
Scientific theories are usually articulated in terms of variables influencing one another.  Therefore,  it is natural,  in a machine learning (ML) interpretability context,  to seek to explain a prediction as the sum of contributions from predictors included in the model.   We refer to this as the primal route to interpretation.  Its key bottleneck is the number of features. Even in the simplest linear regression,  the partial derivative of the predictand with respect to a predictor becomes nearly meaningless in a system featuring 150 cross-correlated variables.  Things do not improve with nonlinear methods.  There exist many ways to address the high-dimensionality interpretation issue,  and all hinge on reinstating some form of sparsity in the \textit{covariate space},  through selection or the construction of latent states. 

 In macroeconomic forecasting, there is something that requires no further sparsification efforts:  the number of training observations.  While the statistical costs of short time series are well documented, this paper highlights an unexpected benefit. It explores the possibility of interpreting any out-of-sample (OOS) forecast through its \textit{dual} representation as the sum of contributions from each training data point.  Interpretation relies on pairwise observation proximity as perceived by the machine learning model, which is quantified by the corresponding \textit{training observation weights}. We call this the {dual route} to interpretation. Clearly, the sparse and ordered nature of the macroeconomic data becomes an advantage.   It is easy to (i) visualize the importance of each historical data point for the forecast,  (ii)  assess  the similarity between current and past economic conditions based on narrative evidence,  and (iii) possibly judge on the credibility of selected economic events. Moreover, viewing proximity scores as \textit{data portfolio weights}, the analysis can be complemented with forecast summary statistics inspired by portfolio construction in finance. This includes forecast concentration, short position, turnover, and leverage.

 \vskip 0.15cm
{\noindent \sc \textbf{Duality: Correlation- and Proximity-based Forecasting.}}  We introduce a general and novel approach to interpret predictions made by macroeconomic forecasting models through data portfolio weights and historical contribution of observations.  We show that those weights have an intuitive proximity-based interpretation through the lesser-known dual solution to least squares-based problems. This provides economic analysts with a gateway to explain machine learning and econometric forecasts to policymakers as quantifiable combinations of historical analogies. 

For estimators that are linear combinations of the target training observations, such as (kernel) ridge regression, the data portfolio weights, can be recovered directly from the primal or dual solution without further calculations.  In fact, for linear models,  the dual solution is simply a tool to reinterpret traditional correlation-based forecasts as proximity-based forecasts, leveraging the matrix inversion lemma.  While both approaches yield identical forecasts, the former is understood through regression coefficients, whereas the latter is interpreted via proximity coefficients.  For models formulated in the dual space, such as kernel ridge regression, only the proximity coefficients are readily available from the estimation process. 

Our approach embraces the implications of the representer theorem, which asserts that any minimizer of a regularized empirical risk functional over a reproducing kernel Hilbert space (RKHS) can be expressed as a \textit{finite} linear combination of kernel functions evaluated at the training data points \citep{kimeldorf1971some, scholkopf2001generalized}.  While the theorem is typically valued for its computational advantages, in this work, we leverage its interpretability potential, particularly when the number of training observations is small and has a predefined temporal structure. 

 \vskip 0.15cm
{\noindent \sc \textbf{Neural Networks and Tree Ensembles.}} Some models, however, do not hand in their proximity matrix as easily.  Such is the important case of neural networks.  We show that for neural networks with a linear output layer—which is the most common configuration for regression problems—it is possible to obtain the proximity weights exactly in the same manner as we do for ridge regression. The key insight is that, when the previous layers are held constant, optimizing the final layer reduces to a regularized linear regression over an expanded set of features generated by the network. As a result, identical forecasts can be produced by an auxiliary ridge regression using the outputs of the second-to-last neurons as regressors. The same calculations applied in the ridge regression case can then be used to derive proximity weights and contributions. 

Tree-based models, such as random forest and boosted trees, can also be interpreted similarly by utilizing standard package outputs and performing accounting operations across the ensemble. This is not surprising either, given that decision trees function as supervised clustering algorithms based on observable predictors—their design is inherently motivated by proximity information.  For those, we can rely on algorithms previously put forward in the literature \citep{rosaler2023,koster2024simplifying,geertsema2023}.

Obtaining forecast contributions is straightforward in \textit{classification} problems for random forest, but models with a sigmoid wrapper function, like (kernel) logistic regression and neural networks, require extra steps. We demonstrate how this can be achieved using an auxiliary dual kernel logistic regression.

We now survey three related literature strands and discuss how to locate this paper's contributions within the existing landscape.

\vskip 0.15cm
{\noindent \sc \textbf{Proximity-based Econometrics.}}  Several studies \citep{dendramis2020similarity, foroni2022forecasting, guerroon2023macroeconomic, lin2021minimizing, lundquist2024volatility} have explored how similarity-based methods may improve forecasts -- sometimes by adjusting linear model coefficients based on the resemblance between current economic conditions and previous historical episodes. Others have designed neural network architectures to directly learn the proximity function between observations, rather than the standard predictive function \citep{martinez2022}. Our paper contributes to this literature by pointing out that there is no fundamental distinction between correlation- and proximity-based forecasting, and that proximity-based interpretation is available for any machine learning model, whether originally designed as such or not. In other words, we do not need to write a nearest-neighbors model to get a nearest-neighbors interpretation.

While the correlation-based paradigm is evidently dominant in macroeconometrics, there are notable contributions which can be grouped under the wide umbrella of "proximity-based" approaches.  For instance, clustering approaches have been used to group historical economic crises \citep{sayek2014financial,raihan2017predicting}.  Some causal studies using propensity scores also qualify as such \citep{angrist2011causal}. \cite{angrist2018semiparametric} extend a matching estimator to estimate impulse response functions through local projections, with propensity scores serving as one approach to define proximity between observations for matching. The matching literature, in fact, utilizes a range of distance measures, with more advanced techniques incorporating kernels similar to those used in kernel ridge regression and support vector machines.

\vskip 0.15cm
{\noindent \sc \textbf{Headwinds for Interpretable Macroeconomic Forecasting.}} The most common approach for interpretability tasks is to work in the primal space,  i.e., using feature-based explanations.  Early suggestions for nonparametric ML models were variable (permutation-based) importance and partial difference plots, more recent ones include  SHapley Additive exPlanations (SHAP) \citep{lundberg2017shap} and Local Interpretable Model-Agnostic Explanations (LIME) \citep{ribeiro2016}. SHAP and LIME both offer a linear decomposition of predictions in terms of each regressor's contribution, and thus become less informative in the absence of regressors' sparsity.  
Given that in macroeconomic datasets, like FRED-MD or FRED-QD, sparsity does not naturally arise from the modeling environment, it is frequent to see dense importance rankings of predictors with wide implicit confidence intervals \citep{medeiros2019,longo2022neural,klieber2024dimred}.

Therefore,  sparsity of any kind,  is at the heart of ML and econometric interpretability.  There are a few options to restore it. First, we can handpick relevant variables \citep{buckmann2022interpretable}, or group features into categories based on prior knowledge \citep{anatomy2}. Second, we can deploy variable selection tools in a pre-processing step \citep{borup2020targeting,timpaper}. However,  with a large set of heavily correlated variables, problems arise, and although many refinements exist \citep{medeiros2016l1,babii2022machine,huber2021inducing},  they often hinge on linearity, and more importantly,  all rely on the credibility of the sparsity assumption when modeling macroeconomic data \citep{illusion}.

A well-known \textit{dense} alternative are factor models,   where sparsity is restored in a latent space.  Factors are typically mutually uncorrelated, a significant advantage.  However,  their interpretation is viscous and dependent on a generous set of untestable assumptions \citep{boivin2006factors,kaufmann2017identifying,despois2023identifying}.  Efforts to preserve interpretability include hybrid approaches that build a priori structure loosely inspired from economic theory on the otherwise flexible nonlinear learning algorithm  \citep{MRFjae,HNN}.  While using a low-dimensional latent space, these models differ from factor models by relying on supervised estimation through an auxiliary \textit{observable} objective (e.g., forecasting inflation).  This strategy, however, requires some a priori structure from modeling tradition, which becomes challenging when considering nontraditional targets, or using features not yet internalized in macroeconomic models.

Thus,  within the ML-based macroeconomic forecasting space,  we contribute by providing a generally applicable interpretation tool for models that are not inherently interpretable, nor lend themselves easily to the use of traditional post-hoc methods in the primal space.  And for those that do,  it provides a complementary explanation bringing in unique information of its own.

\vskip 0.15cm
{\noindent \sc \textbf{Instance-based Interpretability in Machine Learning.}} Several papers in the ML literature have noted that evaluating the importance of individual training observations in driving  out-of-sample predictions can be interesting in its own right.  Key references include \cite{ghorbani2019datashap}, \cite{jia2019towards}, \cite{koh2017understanding}, \cite{pruthi2020estimating}, and \cite{cho2020interpretation}.  For instance,  \cite{ghorbani2019datashap}  propose to fairly attribute value to data points in deep learning pipelines. Their focus is not on interpretability per se but rather on assessing how overall model performance (e.g., test set accuracy) degrades when, for instance, the top 10\% most important training examples are removed compared to the bottom 10\%.  Recent work has also focused on individual algorithms, such as random forest \citep{rosaler2023,koster2024simplifying} and Gradient Boosting Machines \citep{geertsema2023}. While the concept of proximity and the connection to adaptive nearest neighbors in tree-based models have been established for some time \citep{lin2006random, rhodes2023, jeyapaulraj2022rfproximity, li2024quantile}, their application for interpretability purposes is relatively new.

A key challenge in this area is that, unlike macroeconomic time series, cross-sectional data lacks a natural arrangement of observations lending itself nicely to visualization. Moreover, with large datasets,  proximity coefficients abound. Thus, investigating these individually offers a low return on investment. While the time series equivalent is not immune to the need of additional thinking,  narrative knowledge about historical events and the possibility of calculating moving averages greatly simplifies the excavation process.  In contrast to recent contributions, we provide a unified treatment of all major classes of ML models used in macroeconomic forecasting through the dual solution and the concept of data portfolio weights. We also introduce a computationally economical approach to do so for neural networks,  addressing an open problem in the literature. 

\vskip 0.15cm
{\noindent \sc \textbf{Summary of Empirical Results.}} We conduct a machine learning-based macroeconomic forecasting exercise, decomposing key forecasts from the post-pandemic era and the Great Recession. The model suite includes ridge regression, factor-augmented autoregression (FAAR), kernel ridge regression (KRR), random forest (RF), boosted trees, a standard feed-forward neural network (NN), and \cite{HNN}'s Hemisphere Neural Network (HNN).

Our first application focuses on predicting the post-pandemic inflation surge, where many models struggled to (i) capture the impact of the initial Covid-19 shock and (ii) anticipate the rise in 2021-2022.  As expected, ML models have little guidance on where to look when forecasting headline inflation for 2020Q3, following the historically unprecedented 2020Q2. All models produced forecasts below the realized 4.5\%, with some predicting as low as -5\%.  We find that the understandable forecast error comes from highly concentrated forecasts allocating a disproportionate weight to the Great Recession--particularly 2008Q4, similarly marked by a sharp drop in oil prices.   Some models also incorrectly label the extreme conditions of early 2020 as the polar opposite of the late 1970s inflation spike, pushing forecasts further into deflation territory. Withdrawing such inferences brings forecasts much closer to the final reading.


While our models all agree on the importance of the 1970s when predicting the peak in 2022Q2, there is much more dispersion in historical inferences when forecasting the contentious 2021Q1-Q2 period. We find that NN, ridge regression, and boosting align themselves with the transitory inflation team, producing forecasts marginally above the 2\% target by leveraging  perceived proximities to pre-pandemic years. KRR and HNN differ sharply by indicating that business may not be back to usual. KRR significantly upweights the period spanning from 1974 to 1986. For 2021Q1, HNN is still ambiguous by giving significant weight to one of the 1970s inflation spikes, but not two.   By 2021Q2, much of that ambiguity is resolved, with HNN paying much  of its attention to \textit{both} of the 1970s inflation run-ups. Therefore, an analyst having access to this information knows as early as mid-2021 that there is quantitative backing to believe that (high inflation) history is about to repeat itself. 

The second application examines forecasting GDP growth and the unemployment rate during the Great Recession. {\color{black} As its name implies, ML models may have limited historical precedents that are “Great” enough to draw inferences from.} Both the factor-augmented autoregression (FAAR) and NN forecast accurately the mildly negative value for 2008Q1. NN, in particular, associates late 2007 conditions with \textit{all} recessions, emphasizing the 1974 recession. As quantified by our forecast leverage metric, we find that NN amplifies such narratives to procure a significantly negative forecast for 2008Q4.  We find that unemployment results at various horizons for the Great Recession peak of 2009Q1 mostly align with those of GDP growth. 

We conclude with post-pandemic predictions for GDP growth and recession probabilities. Recession fears surged in 2022 after rapid rate hikes by the Federal Reserve. These fears resurfaced in the second half of 2024, with highly scrutinized indicators like the Sahm rule pointing to elevated risks.  Once again, we find models divided in their assessment of historical proximities,  echoing that of professional forecasters. While some models are less pessimistic, NN and RF predict a mild contraction for 2025Q1, heavily weighting the financial crisis, the early 1990s recession, and the second of the twin recessions. Predicting recessions directly through classification problems also shows RF finding a strong proximity of 2024 conditions to the Great Financial Crisis (GFC) and past rapid monetary tightening cycles, resulting in a probability of nearly 70\%. Although there is reason for concern, it is tempered by the fact that these same models (excluding classification RF)  jumped the gun in 2022, predicting a significant contraction due to their strong emphasis on prior tightening cycles. 

\vskip 0.15cm
{\noindent \sc \textbf{Outline.}} This paper is organized as follows.  Section \ref{sec:dual} motivates the dual approach to interpretation, shows how to recover data portfolio weights for various machine learning models, and introduces the forecast summary statistics toolbox. Section \ref{sec:emp} presents empirical results for four applications. Section \ref{sec:con} concludes.

\section{The Dual Route to Interpretation}\label{sec:dual}
 
In this section, we first motivate the use of data portfolio weights for interpretation through the dual solution of convex optimization problems. Next, we demonstrate how to recover these weights and how to seamlessly obtain the contribution of each training observation to a hold-out sample prediction.

We illustrate this process for five canonical classes of machine learning models. We start with the simplest case, ridge regression, and discuss how correlation-based forecasts have a dual interpretation as proximity-based forecasts. Then, we extend these insights to obtain data portfolio weights in nonlinear models, such as kernel ridge regression and neural networks, which can also be formulated through least squares with (implicit) basis expansions. Finally, we explore how to obtain weights in greedily-optimized models, such as random forest and boosting.  Practical extensions of these ideas to classification problems are also presented. Lastly, we discuss derivative metrics of interest that can be constructed from the time series of proximity weights, such as forecast concentration and short position.

The notation used throughout is $i \in \{1,\dots, N\}$ for denoting training observations and $j$ denoting one from the test set.  Thus,  in a time series context,  $i$ would be $t$,  $N$ could be $T$ (information up to today) and $j$ could be $T+1$ for a one step ahead out-of-sample forecast.

 

%

\subsection{Ridge Regression}\label{sec:rr}

We begin with the ridge regression case, which offers a closed-form solution for coefficients and, as we will see,  data portfolio weights.  The ridge regression (RR) coefficients are obtained via:
\begin{equation}\label{ridgepro}
\hat{\boldsymbol{\beta}}= \arg\min_{\boldsymbol{\beta}} \sum_{i=1}^{N} ( y_i - \boldsymbol{\beta}'{X}_{i})^2 + \lambda ||\boldsymbol{\beta}||_2^2 
\end{equation}
where \( || \cdot ||_2 \) is the \( l_2 \) norm. The latter is equivalently \( \sum_{p=1}^{P} \beta_p^2 \) in summation notation, where $P$ is the number of columns of $\boldsymbol{X}$.  We assume throughout for simplicity that both $\boldsymbol{X}$ and $\boldsymbol{y}$ have been standardized before estimation. 

The solution to the ridge problem can be expressed in the following two forms. The first is well known and is a simple generalization of the ordinary least squares formula using covariances:
\[
\hat{\boldsymbol{\beta}}= (\boldsymbol{X}'\boldsymbol{X} + \lambda {I}_P)^{-1} \boldsymbol{X}'\boldsymbol{y} \, .  \tag{Primal Solution}
\]
What is less known is that there is a numerically equivalent solution
\[
\hat{\boldsymbol{\beta}}= \boldsymbol{X}' (\boldsymbol{X}\boldsymbol{X}' + \lambda {I}_N)^{-1} \boldsymbol{y} \tag{Dual Solution}
\]
obtained from solving the convex problem in \eqref{ridgepro} through its dual representation.  Alternatively, the primal solution can be rewritten directly as the dual one (and vice versa) using the matrix inversion lemma \citep{boyd2004convex}. The dual representation,  which enables things such as kernel ridge regression,  can sometimes be exploited to significantly reduce the computational burden in linear models \citep{MRFjae}.  Indeed, when $P$ greatly exceeds $N$, then the dual solution is far more economical as the inner product approach implies the inversion of an $N \times N$ matrix rather than a $P \times P$ matrix, as is necessary for the primal solution.  In this paper,  we notice that the dual approach can also be used to simplify \textit{interpretation},  especially when $P$ is large and $N$ is not.  


\vskip 0.25cm
{\noindent \sc \textbf{Some Intuition for the Dual Solution}}.  Before digging into why and how the dual solution can bolster interpretability in machine learning models, it is worthwhile to get a sense of where it comes from.   First, the primal problem \eqref{ridgepro} can be equivalently formulated as  
\begin{equation} \label{constrained}
\arg \min_{\boldsymbol{\beta},\boldsymbol{r}} \frac{1}{2} \left( \boldsymbol{r}' \boldsymbol{r} + \lambda \boldsymbol{\beta}' \boldsymbol{\beta} \right) \quad \text{subject to} \quad \boldsymbol{r} = \boldsymbol{X}\boldsymbol{\beta} - \boldsymbol{y} 
\end{equation}
as observed in \cite{saunders1998ridge} and others.  Its Lagrangian is
\begin{equation} \label{lagrange}
L(\boldsymbol{\beta}, \boldsymbol{r}, \boldsymbol{a}) = \frac{1}{2} \boldsymbol{r}' \boldsymbol{r} + \frac{\lambda}{2} \boldsymbol{\beta}' \boldsymbol{\beta} + \boldsymbol{a}' (\boldsymbol{r} - \boldsymbol{X}\boldsymbol{\beta} + \boldsymbol{y}) ,
\end{equation}
where $\boldsymbol{a} \in {\rm I\!R}^N$ is a vector of Lagrange multipliers.  Setting derivatives with respect to the primal variables $(\boldsymbol{\beta}, \boldsymbol{r})$ to zero, we obtain from first order conditions that the solution should satisfy $\boldsymbol{\beta} = \frac{1}{\lambda} \boldsymbol{X}' \boldsymbol{a} \quad \text{and} \quad \boldsymbol{r} = -\boldsymbol{a}$.  Making these substitutions to eliminate \(\boldsymbol{r}\) and \(\boldsymbol{\beta}\) gives the dual problem 
\begin{equation} \label{dual}
\arg \min_{\boldsymbol{a}} -\frac{1}{2} \boldsymbol{a}' \boldsymbol{a} - \frac{1}{2\lambda} (\boldsymbol{X} \boldsymbol{a})' (\boldsymbol{X} \boldsymbol{a}) + \boldsymbol{a}' \boldsymbol{y},
\end{equation}
where everything is expressed in terms of \(\boldsymbol{a}\) rather than \(\boldsymbol{\beta}\).  From the first order conditions, particularly $\boldsymbol{r} = -\boldsymbol{a}$, it is easy to see what this dichotomy implies. In traditional statistical practice, one solves for coefficients and deducts residuals \(\boldsymbol{r}\) from what those imply in terms of fitted values. Alternatively, one could solve for residuals (subject to some constraints) and deduct corresponding coefficients.  While this constitutes a needless detour in low-dimensional settings ($P<<N$), it becomes the most straightforward route when $P>>N$.

Moving towards a proximity-based representation, we can reparametrize, defining \(\boldsymbol{\alpha} = \frac{1}{\lambda} \boldsymbol{a}\)  and directly use  knowledge about the dual solution (\( \boldsymbol{\beta}= \boldsymbol{X}' \boldsymbol{\alpha}\)) in the primal problem \eqref{ridgepro}.  We obtain
\begin{align*}
\min_{\boldsymbol{\alpha}}\left(\boldsymbol{y}-\boldsymbol{K\alpha}\right)'\left(\boldsymbol{y}-\boldsymbol{K\alpha}\right)+\lambda \boldsymbol{\alpha'K\alpha},
\end{align*}
where $\boldsymbol{K}=\boldsymbol{XX}'$ is a  kernel matrix, or a matrix of proximity scores, here defined by the inner products in a Euclidean space. Solving for $\boldsymbol{\alpha}$ gives $\hat{\boldsymbol{\alpha}} =  (\boldsymbol{K} + \lambda {I}_N)^{-1}\boldsymbol{y}$ and the dual formula for \(\hat{\boldsymbol{\beta}}\) is obtained through \( \hat{\boldsymbol{\beta}}= \boldsymbol{X}' \hat{\boldsymbol{\alpha}}\).


\vskip 0.25cm

{\noindent \sc \textbf{The Data Portfolio}}.  We now present our main object of interest for interpretation, data portfolio weights $\boldsymbol{w}_j $, and provide an intuitive explanation of their formula using the dual solution. First, define ${K}_{j}  = X_j  \boldsymbol{X}'$ as the $1 \times N$ vector of proximity scores  of test observation $j$ with respect to each of the training observations.  A prediction for an out-of-sample observation \(j\) can be obtained from two numerically equivalent formulas 
\begin{align}
\hat{y}_j &= X_j \hat{\boldsymbol{\beta}}= X_j (\boldsymbol{X}'\boldsymbol{X} + \lambda {I}_P)^{-1} \boldsymbol{X}'\boldsymbol{y} \tag{\text{Correlation-Based Prediction}} \\
         \hat{y}_j &=  \hspace{0.1em} K_j \hat{\boldsymbol{\alpha}}  = X_j  \boldsymbol{X}' (\boldsymbol{X}\boldsymbol{X}' + \lambda {I}_N)^{-1} \boldsymbol{y}  \tag{\text{Proximity-Based Prediction}}
\end{align}
offering different decompositions for $\hat{y}_j$.  Significant efforts are usually given to analyzing $\hat{\boldsymbol{\beta}}$ directly, and their analog through Shapley Values in nonparametric models. Yet, for a given prediction,  the $N \times 1 $ vector $\boldsymbol{w}_j  \equiv  X_j  \boldsymbol{X}' (\boldsymbol{X}\boldsymbol{X}' + \lambda {I}_N)^{-1}$ can be revealing in its own right,  especially if it is shorter in length than  $\hat{\boldsymbol{\beta}}$.  In fact, there is no conceptual distinction between correlation-based or proximity-based, as one can be rewritten as the other without any additional assumption.   By virtue of RR being a linear estimator,  whatever route we wish to take to obtain the solution,  we have that 
\begin{align}
\hat{y}_j = \boldsymbol{w}_j \boldsymbol{y} \quad \quad \forall \enskip j \in \text{Test Sample} \, . 
\end{align}
The primal meaning of the data portfolio weights $\boldsymbol{w}_j$ is that of a regularized "out-of-sample projection matrix". The dual route 
\begin{align}\label{wk}
 \boldsymbol{w}_j  = {K}_{j}  (\boldsymbol{K}  + \lambda {I}_N)^{-1} 
\end{align}
also provides an exact breakdown of $\hat{y}_j$ into training data points \textit{and} comes with an appealing and intuitive explanation.  The entries of \( {K}_j \) have higher values when the corresponding values in \( {X}_j \) are more similar to those in \( {X}_i \) for a given \( i \). In this linear model, this notion of distance is  the Euclidean inner product \( {X}_j' {X}_i \). $\boldsymbol{K}$ is the in-sample equivalent of the same concept: it is the square matrix of proximity weights for all possible pairs of training observations.  As such, the "denominator" $(\boldsymbol{X}\boldsymbol{X}' + \lambda {I}_N)^{-1}$ is the regularized proximity matrix of all $i$'s with respect to each other.   With $\lambda \rightarrow \infty$,  $\hat{\boldsymbol{\beta}}\rightarrow 0$, but also $(\boldsymbol{X}\boldsymbol{X}' + \lambda {I}_N)^{-1} \rightarrow \mathbf{0}$, nullifying any proximity information in the numerator, and reporting what is here the unconditional mean of 0 (for pre-standardized data).  Thus,  ${w}_{ji}$ is a normalized (and regularized) measure of proximity between the out-of-sample observation $j$ and the in-sample one $i$.  If economic conditions characterized by  ${X}_j$ are similar  to those of  ${X}_i$, $y_i$ gets upweighted in $\hat{y}_j $. 


Note that the dual representation of forecasts extends to linear models even in the absence of ridge regularization (i.e., $\lambda = 0$). This is because the ordinary least squares (OLS) estimator can be expressed as:
\[
\hat{\boldsymbol{\beta}}_{\text{OLS}} = (\boldsymbol{X}'\boldsymbol{X})^{-1} \boldsymbol{X}'\boldsymbol{y} = \boldsymbol{X}' (\boldsymbol{X} \boldsymbol{X}')^{+} \boldsymbol{y} \, ,
\]
leveraging the properties of the generalized inverse, also known as the Moore-Penrose pseudoinverse, denoted by $(\boldsymbol{X}\boldsymbol{X}’)^{+}$. The generalized inverse satisfies specific algebraic conditions that ensure a solution even when the matrix $\boldsymbol{X}$ has less predictors than observations ($P < N$). Consequently, forecasting equations estimated via OLS, such as those used in factor-augmented autoregressions, inherently admit a proximity interpretation. 


\vskip 0.25cm

{\noindent \sc \textbf{A Simple Case}}.  These insights are easy to visualize in the simple case where we have one training observation ($\{y_1 , \, X_1 \}$) and one test observation  ($\{y_2 , \, X_2 \}$).  In this environment, \eqref{wk} reduces to
\begin{align}\label{w2eq}
w_2 = \frac{\langle {X}_2, {X}_1 \rangle}{\langle {X}_1, {X}_1 \rangle + \lambda}   \, .
\end{align}
Setting $\lambda=0$,  it is easy to see that the denominator is normalizing the units of the numerator by dividing by the measure of maximal in-sample proximity (${X}_1$ with itself).  Specifically, if $X_1 = X_2$, we get $w_2 = 1$, and consequently, $\hat{y}_2 = y_1$. A similar logic applies for very dissimilar $X_1$ and $X_2$, for which $w_2$ can be negative and will be equal to $-1$ if $X_2 = -X_1$. The effect of $\lambda$ is to shrink $w_2$ towards the conservative value of 0 (i.e.,  ${X}_1$ and ${X}_2$ are not similar nor very dissimilar) by "artificially" inflating the notion of maximal in-sample proximity in \eqref{w2eq}.  

To gain sharper geometric insights, \eqref{w2eq} can be rearranged for the $\lambda = 0$ case as
\begin{align}\label{cos}
w_2 = \cos(\gamma) \frac{ || X_2 ||_2}{ || X_1 ||_2} 
\end{align}
where $\gamma$ is the angle (in degrees) between the vectors $X_1$ and $X_2$. The scaling factor $\frac{\|X_2\|_2}{\|X_1\|_2} = \sqrt{\sfrac{\langle {X}_2, {X}_2 \rangle}{\langle {X}_1, {X}_1 \rangle}}$ represents the ratio of the $l_2$ norms of the out-of-sample feature vector to that of the in-sample one.  By construction, $\cos(\gamma) \in [-1, 1]$, with $\cos(\gamma) = 1$ indicating perfect similarity between $X_1$ and $X_2$, $\cos(\gamma) = -1$ indicating perfect dissimilarity (opposite directions), and $\cos(\gamma) = 0$ indicating no similarity (orthogonality). Thus, the weight attributed to $y_1$ in predicting $y_2$ is proportional to how close $X_1$ and $X_2$ are in $\mathbb{R}^P$ based on their Euclidean distance. Bringing back $\lambda$ in \eqref{cos} implies using an artificially inflated $l_2$ norm for $X_1$, i.e., $\tau \cdot  \langle {X}_1, {X}_1 \rangle$ with $\tau = 1 + \frac{\lambda}{\langle {X}_1, {X}_1 \rangle}$  for both the definition of the angle and the denominator of the scaling factor. 

The scaling factor $\frac{\|X_2\|_2}{\|X_1\|_2}$ emphasizes that the raw $w_2$ is not scale-invariant. While the normalized proximity $\cos(\gamma)$ lies within $[-1, 1]$, this scaling can either amplify or compress its effect depending on the relative magnitudes of $\|X_2\|_2$ and $\|X_1\|_2$. For instance, if $X_2$ exhibits significantly greater dispersion than $X_1$ -- as might occur when using a model trained on pre-2020 data to forecast in 2020 -- we can observe $w_2 > 1$, reflecting both positive proximity and extrapolation. Depending on the objective of the analysis, it can sometimes be more insightful to examine raw weights, as they capture the combined effects of both channels and ultimately produce the forecast itself. Alternatively, normalized weights may be more suitable for comparing perceived proximities across forecasts made at different dates, since $\|X_2\|_2$ varies while $\|X_1\|_2$ remains fixed (assuming an unchanged training sample).

\vskip 0.25cm

{\noindent \sc \textbf{Remarks}}.  The equivalence between proximity- and correlation-based forecasting, evident even in the simplest linear models, carries its own insights. It suggests that if a model is sufficiently flexible in its original correlation-based formulation, there should be little room for improvement through post-estimation proximity-based adjustments (as in, e.g., \cite{dendramis2020similarity} and \cite{foroni2022forecasting}). This, of course, does not imply that such adjustments should always be avoided—some modeling aspects may be more practically implemented in one space than the other. Rather, it means that there is no fundamental distinction between refining the model directly in the primal space and enhancing it with proximity-inspired alterations.


Viewing $\boldsymbol{w}_j$ as portfolio weights inspires the development of forecast summary statistics, as discussed in Section \ref{sec:derivatives}. This perspective also bridges ML forecasts with other areas of econometrics-based forecasting literature. A notable example is forecast combinations, where a meta-forecaster constructs a linear aggregation of forecasts, such as those from professional forecasters, to achieve better predictive performance  \citep{wang2023forecast}. Through $\boldsymbol{w}_j$, ML models can be interpreted as a time-varying forecast combination scheme, where the individual forecasts are the training realizations of the dependent variable, and the time-varying weights (varying with $j$) are proportional to a proximity score between $j$ and $i$.

\vskip 0.15cm
{\noindent \sc \textbf{Coefficients \textit{vs. }Contributions}}.  As in the primal space, one can analyze either coefficients or contributions. Although \(\beta_p\) is the natural object to focus on for the global interpretation of a fitted model, it is equally natural to examine \(c_{jp}\) when interpreting a single prediction \(\hat{y}_j\), since, by construction, \(\hat{y}_j = \sum_{p=1}^{P} c_{jp}\) where \(c_{jp} = X_{j,p} \hat{\beta}_p\). Moreover, post-hoc interpretability methods such as Shapley Values deliver contributions, even in linear models.

Conversely, in the dual approach, \(w_{ji}\) represents the attention given to each observation in \(\boldsymbol{y}\), and the contributions \(c_{ji} = w_{ji}y_i\) provide an exact linear decomposition of \(\hat{y}_j\) (i.e., \(\hat{y}_j = \sum_{i=1}^{N} c_{ji}\)). In some cases, $w_{ji}$, may have a high value and showcase high concentration on a particular time period, but if its associated $y_i$ is close to the unconditional mean (e.g., during periods like the Great Moderation), its effect on the final forecast will be negligible. Therefore,  while $c_{ji}$ is the product of two components that we may want to analyze separately, its synthesis of both $w_{ji}$ and $y_i$ graphically elicits relevant information of its own.


\subsection{Kernel Ridge Regression}\label{sec:krr}

One could  argue that looking at $\boldsymbol{w}$ is not overly exciting for a model where $\hat{\boldsymbol{\beta}}$'s are available and $P$ is {small}.  However,  in a \textit{kernel} ridge regression (KRR) context,  looking at $\boldsymbol{w}$ is all the more natural given that the formulation of the optimization problem and the introduction of nonlinearities are already done through the dual path.\footnote{KRR is also known under the name Gaussian Processes (GPs),  particularly in Bayesian statistics.  Recent applications of GPs in macroeconometrics include \cite{clark2024forecasting,hauzenberger2024gaussian,hauzenberger2024nowcasting}.} The kernel trick -- which is the catalyzer of nonlinearities in KRR,  support vector machines (for classification),  and support vector regression (regression with an alternative loss function) -- can be understood as an \textit{implicit} basis expansion technique.  Presume we wish to increase the sophistication of our prediction function by now considering $\Phi({X_i}) \in  \mathbb{R}^{\tilde{P} \times N}$ as feature matrix (with $\tilde{P}>P$).  $\Phi({X_i})$ is an expanded set of regressors constructed from the original inputs.  In the $P=2$ case,  a possible expansion is
\[
\Phi(x_i) = \left[ 1\quad 2x_{i,1}\quad 2x_{i,2}\quad x_{i,1}^2\quad \sqrt{2}x_{i,1}x_{i,2}\quad x_{i,2}^2 \right] 
\]
and the prediction for $y_j$ would be generated through
\[
\begin{aligned}
\hat{y}_j & =\Phi({X}_j) \hat{\boldsymbol{\beta}} \\
& =\Phi({X}_j)\left(\Phi(\boldsymbol{X})^{\prime} \Phi(\boldsymbol{X})+\lambda {I}_{\tilde{P}}\right)^{-1}  \Phi(\boldsymbol{X})^{\prime} \boldsymbol{y} \\
& = \Phi({X}_j) \Phi(\boldsymbol{X})^{\prime}\left(\Phi(\boldsymbol{X}) \Phi(\boldsymbol{X})^{\prime}+\lambda {I}_{N}\right)^{-1}  \boldsymbol{y} \\
&=  \mathcal{K}(X_j,  \boldsymbol{X}) ( \mathcal{K}( \boldsymbol{X},  \boldsymbol{X}) + \lambda {{I}_N})^{-1} \boldsymbol{y} \\
& = \underbrace{{K}_j ( \mathbf{K} + \lambda {I}_N)^{-1}}_{\boldsymbol{w}_j}  \boldsymbol{y} \, .
\end{aligned}
\]
Therefore,  all we need to generate forecasts (and obtain $\boldsymbol{w}_j $) from this nonlinear model in $\boldsymbol{X}$ is the $\mathbf{K}$ matrix and its associated reproducing kernel function $\mathcal{K}$.  The matrix  $\mathbf{K} \in  \mathbb{R}^{N \times N}$ reproduces the inner product one would obtain from a possibly infinite-dimensional $\Phi({X_i}) $, and thereby its predictions.  Thus,  there is no need to formulate the $\Phi()$ function as $\mathcal{K}$ introduces nonlinearities directly by altering the notion of proximity between training data points.

There are many possible kernel functions, such as that generating quadratic polynomials \(\mathcal{K}\left(X_{i}, X_{j}\right)=\left(1+X_{i}^{\prime} X_{j}\right)^{2}\). Other options include the Gaussian kernel (\(\mathcal{K}(X_{i}, X_{j}) = \exp\left(-\sfrac{\|X_{i} - X_{j}\|^2}{2\sigma^2}\right)\)) and the Laplacian one (\(\mathcal{K}(X_{i}, X_{j}) = \exp\left(-\sfrac{\|X_{i} - X_{j}\|}{\sigma}\right)\)).  \textcolor{black}{Note that,} ridge regression is a special case of KRR where the kernel is linear (i.e., \(\mathcal{K}(X_{i}, X_{j}) = X_{i}^\prime X_{j}\)).  While in limit theories the choice of \(\mathcal{K}\) does not matter to reproduce relevant nonlinearities, this choice, along with that of tuning parameters \(\lambda\) and \(\sigma\), can be quite influential in short samples, such as those at the disposal of macroeconomic forecasters.   It is easy to see why:  \(\mathcal{K}\) defines the notion of distance between $i$'s and $j$ which will ultimately be reflected in $\boldsymbol{w}_j $. Moreover, the bandwidth hyperparameter $\sigma$ regulates how local or global the kernel is, which has a direct influence on the concentration (or diversification) of $\boldsymbol{w}_j $.

Understanding the mechanism generating $\hat{y}_j$ in the primal space for KRR requires post-hoc calculations, like partial dependence plots or Shapley Values.  In contrast,  the dual space interpretation is directly available from estimation which can be written in a single line of code using $\mathcal{K}$ functions from \texttt{kernlab} in R or \texttt{scikit-learn} in Python.


 


\subsection{Neural Networks}\label{sec:nn}

%

Neural networks (NN) can be directly interpreted through duality with minimal additional calculations, provided the final layer is linear. This configuration is common in regression problems, regardless of whether the architecture is a plain dense network, a recurrent NN, LSTM, convolutional NN, or even Transformers. 

\begin{figure}
\caption{\normalsize{Neural Network Architecture}} \label{fig:nn2}
\centering
\vspace*{-0.5em}
      \includegraphics[scale=1.2, trim = 0mm 0mm 0mm 0mm, clip]{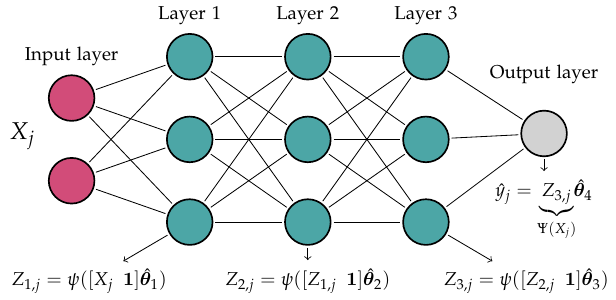}

 \begin{threeparttable}
    \centering
    \vspace*{-1.5em}
    \begin{minipage}{\textwidth}
      \begin{tablenotes}[para,flushleft]
    \setlength{\lineskip}{0.2ex}
    \footnotesize 
  {\textit{Notes}: The figure presents a standard feed-forward neural network with three hidden layers and a linear final layer.}
    \end{tablenotes}
  \end{minipage}
  \end{threeparttable}
\end{figure}

To see this, it is worthwhile to review how simple dense NN's predictions are generated and how the parameters of a given architecture are optimized. This is illustrated in Figure \ref{fig:nn2}. For a test set observation \(j\), \(\hat{y}_j\) can be obtained recursively, moving forward from inputs \(X_j\) towards \(\hat{y}_j\), through
\[
\begin{aligned}
Z_{1,j} &= \psi \left( \left[ X_j \enskip \boldsymbol{1} \right]\hat{\boldsymbol{\theta}}_1 \right) \\
Z_{l,j} &= \psi \left( \left[ Z_{l-1,j} \enskip \boldsymbol{1} \right] \hat{\boldsymbol{\theta}}_l \right), \quad \text{for } 2 \leq l \leq L-1 \\
\hat{y}_j &= Z_{L-1,j} \hat{\boldsymbol{\theta}}_{L}
\end{aligned}
\]
where \(\psi\) denotes a nonlinear activation function and the \(\hat{\boldsymbol{\theta}}_l\)'s are the estimated network's weights at various layers.  The last layer \(L\) can be interpreted as a \textit{linear prediction} with generated features  $Z_{L-1,j}  \equiv \Psi({X}_j)$,  the latter corresponding to the outputs of the last layer \(L\) neurons—a total of \(n_L\) time series where \(n_L\) is the number of neurons in the last layer. In other words, we have,   \(\Psi(\boldsymbol{X}) \in \mathbb{R}^{N \times n_L}\),  our new expanded set of features with which the proximity of \(j\) and \(i\)'s can be calculated using the Euclidean distance, as we did for ridge regression. Following this observation, out-of-sample predictions can be rewritten using the same logic as before:
\begin{align}\label{eq:nn}
\hat{y}_j & = \Psi({X}_j) \hat{\boldsymbol{\theta}}_L  \\
& \cong \Psi({X}_j)\left(\Psi(\boldsymbol{X})^{\prime} \Psi(\boldsymbol{X}) + \lambda {I}_{n_L}\right)^{-1} \Psi(\boldsymbol{X})^{\prime}  \boldsymbol{y} \label{uo} \\
& = \Psi({X}_j) \Psi(\boldsymbol{X})^{\prime}\left(\Psi(\boldsymbol{X}) \Psi(\boldsymbol{X})^{\prime} + \lambda {I}_{N}\right)^{-1}  \boldsymbol{y} \\
& = \underbrace{\mathbf{K}_j ( \mathbf{K} + \lambda {I}_N)^{-1}}_{\boldsymbol{w}_j}  \boldsymbol{y} \, .
\end{align}
Thus, retrieving \(\boldsymbol{w}_j\) only requires \(\Psi({X}_j)\) (directly available in PyTorch or TensorFlow) and applying the same formulas as with ridge regression.  Unlike KRR,  nonlinearities are generated directly in the feature space rather than the proximity space -- but they still imply some non-Euclidean notion of distance between original input vectors. 


Since it is always good practice to ensemble a few optimization runs to integrate out NN weights' initialization uncertainty, we focus on
\[
\hat{y}_j = \frac{1}{B}\sum_{b=1}^B \Psi_b({X}_j) \hat{\boldsymbol{\theta}}_{L,b} = \sum_{b=1}^B  \frac{\boldsymbol{w}_{jb}}{B}  \boldsymbol{y} = \boldsymbol{w}_j \boldsymbol{y}
\]
in our empirical results, where $B$ is the number of constituents in the ensemble.   Note that this representation can also accommodate bootstrap or subsampling schemes as in \cite{HNN}.  These  imply that some of the entries in $\boldsymbol{w}_{jb}$ are set to zero ex-ante.



\subsubsection{On the Accuracy of the Ridge Representation of Neural Networks}

We use "\(\cong\)" to denote the passage between \eqref{eq:nn} and \eqref{uo} because it is not an identity, but rather a very accurate approximation. An identity would require that \(\hat{\boldsymbol{\theta}}_{L} = \left(\Psi(\boldsymbol{X})^{\prime} \Psi(\boldsymbol{X}) + \lambda I_{n_L}\right)^{-1} \Psi(\boldsymbol{X})^{\prime} \boldsymbol{y}\), which could be obtained by running a ridge regression of \(\boldsymbol{y}\) on \(\Psi(\boldsymbol{X})\). While this seems reasonable, the usual forecasting practice with neural networks does not include this additional post-processing step.


However, the key observation is that the first-order conditions derived from gradient descent for the top layers are very well approximated by this "thought experiment" post-processing ridge regression. If the neural network were optimized without early stopping to reach the global minimum, the conditions defining the optimal values for \(\hat{\boldsymbol{\theta}}_{L}\) given fixed \(\hat{\boldsymbol{\theta}}_{1:(L-1)}\) would be \textit{exactly} the same as that of plain least squares (\(\lambda = 0\)).  Considering the problem alternatively optimized through block coordinate descent or an EM algorithm alternating the conditional estimation of \(\boldsymbol{\theta}_{L}\) and \(\boldsymbol{\theta}_{1:(L-1)}\), the first-order conditions for \(\hat{\boldsymbol{\theta}}_{L}\) given \(\hat{\boldsymbol{\theta}}_{1:(L-1)}\) (and thus \(\Psi(\boldsymbol{X})\)) yield a plain least squares closed-form solution. Thus, in this no-regularization scenario, \(\hat{\boldsymbol{y}}\) would correspond exactly to fitted values obtained from regressing \(\boldsymbol{y}\) on \(\Psi(\boldsymbol{X})\), and \eqref{uo} could be granted "=" instead of "\(\cong\)".



However, neural networks are equipped with various regularization mechanisms, including early stopping, double descent, and the use of stochastic batches in optimization \citep{belkin2019reconciling, deeplearning}. This is why we use "\(\cong\)" instead of "\(=\)".  Still,  the intuition developed for the no-regularization case indicates another least squares-like formula is sufficient to back out the desired \(\boldsymbol{w}\) in a broader context. 

While regularization schemes can play various roles in the lower levels of the neural network,  their role for a linear model, which is what the last layer consists of,  is well understood.   Indeed, many authors have linked ridge regularization with gradient descent.   \cite{friedman2004gradient} first observed this connection and  \cite{yao2007early} highlighted that early stopped gradient descent functions as a spectral filter similar to $l_2$ regularization.  Further research for nonlinear models has shown that early-stopped gradient descent performs comparably to explicit $l_2$ regularization when both methods are optimally tuned \citep{bauer2007regularization, raskutti2014early}.  More recently,  \cite{ali2019continuous} study gradient updates in continuous time and compare those to ridge regularization for the linear squares problem,  and find very close correspondence between the risk curves, especially around the optimal point. The ridge connection can also be applied empirically, as demonstrated by \cite{dre}, who approximate neural networks' performance on classic datasets using ridge regressions with randomly generated features.


For our problem of backing out \(\boldsymbol{w}\), this connection is instrumental as it allows us to obtain near-identical predictions from a ridge regression using \(\Psi(\boldsymbol{X})\) as predictors, provided a suitable \(\lambda\) is chosen. The natural strategy is to select \(\lambda\) such that out-of-sample predictions from the original network in \eqref{eq:nn} are closest to those of the auxiliary RR in \eqref{uo} used for interpretation purposes. In practice, the replication accuracy is always higher than 99\%. We found that \(\lambda\) should typically be a small value to ensure that \(\left(\Psi(\boldsymbol{X})^{\prime} \Psi(\boldsymbol{X}) + \lambda I_{n_L}\right)^{-1}\) is not computationally singular, and which approximates well the leftover regularization that is pushing \(\hat{\boldsymbol{\theta}}_{n_L}\) away from the (in-sample) optimal solution. 


Lastly, our "tuning" strategy for $\lambda$ is driven by the specific objective of \textit{replicating} the predictions of the standard network.  However, one could leverage the ridge representation to globally tune the network in post-processing using the simpler cross-validation procedures of ridge regression. In this setup, the dual representation would hold exactly, and whether these predictions could outperform those of a standard network is a question left for future research.

\subsection{Random Forest}\label{sec:rf}

Regression trees of moderate depth are easily interpretable. However, due to their high variance, their performance is  often inferior to that of random forest (RF) and boosted trees, two ensemble methods commonly labeled as black boxes. While this characterization is not entirely accurate in the low-dimensional case or in scenarios where sparsity can be restored through prior macroeconomic knowledge and more sophisticated forest structures \citep{buckmann2022interpretable,MRFjae}, it most certainly is for the "plain" high-dimensional case. The latter is increasingly prevalent with curated databases like those from \cite{mccrackenng} and \cite{mccracken2020fred},  and even more so in the presence of alternative data. Thus, in both low- and high-dimensional settings, interpretation through proximity offers a valuable complement to traditional features-based explanations.




Retrieving \(\boldsymbol{w}_j\) is also fairly straightforward in random forest (RF) since RF's design inherently resembles a supervised clustering problem. The possibility of rewriting RF's predictions as a convex combination of \(\boldsymbol{y}\) was noted long ago, for example, in \cite{lin2006random}. This only requires post-processing of estimation outputs. When predicting for \(j\), we have: 
\[ \hat{y}_j = \frac{1}{B} \sum_{b=1}^B \mathcal{T}_b(X_j) \]
where \(B\) is the number of trees in the RF. Each single tree \(\mathcal{T}_b\) delivers a prediction according to the following rule:
\[
\mathcal{T}_b(X_j) = \frac{1}{\sum_{i=1}^N I\left(i \in \mathcal{P}_b(X_j)\right)}\sum_{i=1}^N y_i I\left(i \in \mathcal{P}_b(X_j)\right) = \sum_{i=1}^N w_{bji} y_i
\]
where \(\mathcal{P}_b\) is the partition implied by the tree and its conditioning information for observation \(j\), and \(w_{bji} = \frac{I\left(i \in \mathcal{P}_b(X_j)\right)}{\sum_{i'=1}^N I\left(i' \in \mathcal{P}_b(X_j)\right)}\). Then, by reordering sums, we get the desired representation:
\[
\hat{y}_j = \frac{1}{B} \sum_{b=1}^B \mathcal{T}_b(X_j) = \frac{1}{B} \sum_{b=1}^B \sum_{i=1}^N w_{bji} y_i = \sum_{i=1}^N \underbrace{\frac{1}{B} \sum_{b=1}^B w_{bji}}_{w_{ji}} y_i = \boldsymbol{w}_j \boldsymbol{y}.
\]
In words, to generate $\boldsymbol{w}_j$ in the RF case, one can follow these steps: determine which leaf observation $j$ falls into for a given tree (based on its $X_j$), identify the corresponding in-sample observations for the leaf and their weights (calculated as \(\sfrac{1}{\text{leaf size}}\)), assign these weights to the relevant in-sample observations (\(w_{bji}\)), and then aggregate these "votes" across all trees in the ensemble.


\vskip 0.15cm
{\noindent \sc \textbf{No Short Position: A Specificity of Random Forest}}.  An important observation emerges when comparing the weights \(\boldsymbol{w}_j\) obtained from tree ensembles to those from linear models,  KRR,  and NN.  In the case of random forest, \({w}_{ji} \geq 0 \, \forall i\) by construction: the \({w}_{ji}\) values are averages of \({w}_{bji}\), which are themselves weighted average weights. From a portfolio construction perspective, this implies a "no short-selling" constraint. For linear models and KRR, there is no such restriction on \({w}_{ji}\), which can take positive or negative values. This absence of such a restriction is not surprising for linear models, as linearity implies symmetry. However, the hard constraint of not utilizing any information from symmetry is quite unique to RF.


The proximity-based view can help in understanding what these restrictions (or their absence) imply when formulating predictions.  If  $X_j$ is in the neighborhood of $X_i$, the observation $y_i$ is featured predominantly in any forecast.  In RF, if $X_j$ and $X_i$ are very dissimilar, we get that ${w}_{ji} \rightarrow 0^+$.  In RR, high dissimilarity rather implies ${w}_{ji} < 0$ through symmetry. Thus, with RF, $y_i$ from very dissimilar times are not featured in $\hat{y}_j$, whereas in linear models, $y_i$ from very dissimilar times will be predominantly used in $\hat{y}_j$ by rotating $y_i$ to $-y_i$.  What happens in KRR or NN depends on the implied basis expansions, but one thing is for sure: there is no such built-in ${w}_{ji} \geq 0$ restriction. More concretely, in any model but RF (or a single tree), if current conditions are very dissimilar from recession ones, the negatives of recession observations can be used to forecast current times. As with short-selling in portfolio construction, this strategy can be effective at reducing variance (through a less concentrated, non-sparse $\boldsymbol{w}_j \boldsymbol{y}$), but it also comes with obvious risks, such as assuming that positive and negative shocks have symmetrical impacts. 

\subsection{Boosting}\label{sec:bt}

The boosting predictive function for a test observation $j$ is 
\begin{align}\label{eq:boost}
\hat{y}_j = \nu \sum_{s=1}^S \mathcal{T}_s(X_j) 
\end{align}
where $S$ is the number of trees,  and $\nu$ a learning rate.  Random forest (RF) and boosted trees (BT) prediction functions appear very similar. Yet, there are significant differences in how trees are generated—RF involves parallel averaging of deep trees, while BT involves sequential addition of shallow ones.  As a result, unlike RF, where the leaves in all trees are local averages of \(\boldsymbol{y}\), boosting's individual tree predictions are averages of ever-changing \textit{pseudo residuals}, not the original \(\boldsymbol{y}\).  Therefore,  a more sophisticated procedure is needed to allocate back contributions to individual $y_i$. 



\vskip 0.15cm

{\noindent \sc \textbf{Obtaining Weights for Regression}}. To uncover these instance weights, we leverage the algorithm proposed in \cite{geertsema2023}. Their AXIL (Additive eXplanations with Instance Loadings) algorithm allows us to write the prediction of BTs as linear combination of in-sample observation weights -- similar to RF. However, given BT's recursive tree-building sequence, the weights retrieval algorithm is also a fairly involved recursion. Therefore, we refer the reader  to \cite{geertsema2023} for a complete exposition of the algorithms and the justifying proofs. 


We note that, unlike in RF, it is possible for BT to exhibit a short position, the magnitude of which depends on the structure of the trees constructed. This occurs because trees indexed by $s > 1$ use pseudo-residuals as targets rather than the original $\boldsymbol{y}$. While short positions cannot occur with respect to pseudo-residuals by construction, they may emerge with respect to the original $\boldsymbol{y}$, which is our focus. This effect becomes evident when computing the overall weights for $\boldsymbol{y}$ in a simplified scenario where $S = 2$ and $\nu = 1$. It is important to note, however, that with the small learning rates typically employed, the short positions observed in our applications are minimal—so minor, in fact, that they are practically comparable to the short positions in RF, which are mechanically zero.

\vskip 0.15cm
{\noindent \sc \textbf{On the Basis Expansion Route}}.   Since \eqref{eq:boost} is a linear predictive equation with homogeneous coefficients \(\boldsymbol{\nu} = \nu \boldsymbol{1}\) given the basis expansions \(Z_{sj} \equiv \mathcal{T}_s(X_j)\), one might wonder if we could apply the same strategy used for neural networks (NN) in Section \ref{sec:nn}. Unfortunately, the logic that justified its use for NN does not directly apply here, as there is nothing in the configuration of standard boosting that implies the first-order conditions of a regularized least squares problem should align with \( \nu \sum_{s=1}^S \mathcal{T}_s(X_j)\). This is because the boosting predictive function is derived through a greedy algorithm, involving a series of local gradient steps based on the outcome of each preceding step. Thus, there is no global, overarching gradient that aligns all trees together.

For equivalence to hold in this context, we would need the least squares solution to produce the homogeneous vector \(\nu \boldsymbol{1}\). While considerable homogeneity can be enforced through \(\lambda\), shrinking coefficients toward a common value of 0, we found that this approximation sometimes provides satisfactory accuracy. However, it does not reach the level of precision achieved by the direct approach described above, which offers an exact decomposition.

\subsection{Classification}\label{sec:class}

The discussion thus far has focused on the prevalent regression cases. But there are interesting macroeconomic forecasting applications which imply binary target variables, such as recession forecasting. We describe here how our proposed apparatus needs to be adapted for this case, which features the additional complexity of the nonlinearity induced by the sigmoid function, bounding predictions to be proper probabilities.

There are three complications. First, we cannot recover weights directly except for the RF case. This is due to the absence of a closed-form solution for RR, KRR, and the auxiliary RR used in neural networks, which are now trained using the log-loss. Indeed, for the logistic versions of such models, we do not have an analytical solution from which we can easily separate $\boldsymbol{y}$ from $\boldsymbol{X}$. Therefore, we focus on contributions which can be backed out directly from the dual solution. 

Second, for the same reasons, (K)RR solutions must now be estimated directly in the dual space because of the absence of a formula mapping regression coefficients into proximity ones. The switch to (kernel) logistic regression implies that we are now solving for $\boldsymbol{\alpha}$ directly, and we do so through gradient descent. In practice, we use a variant of the Adam algorithm to facilitate convergence and ensure the solution path is not overly dependent on the chosen learning rate.


Third, one must be cautious when interpreting cumulative contributions \textit{to probabilities}, as the size of each contribution $c_{ji}$ is not order-invariant. In a logistic regression, the marginal effect of a variable depends on the current prediction level. With slight abuse of notation, we can visualize this through 
\[
c_{ji}^{\text{proba}} \equiv \frac{\partial \hat{\mathbb{P}}(y_j = 1 | 1:i )}{\partial i } = c_{ji}^{\text{log-odds}} \cdot \hat{\mathbb{P}}(y_j = 1 | 1:i ) \cdot (1 - \hat{\mathbb{P}}(y_j = 1 | 1:i )) \, .
\]
This shows that the marginal effect on the probability ($c_{ji}^{\text{proba}}$) scales $c_{ji}^{\text{log-odds}} $ by $\hat{\mathbb{P}}(y_j = 1 | 1:i ) \cdot (1 - \hat{\mathbb{P}}(y_j = 1 | 1:i ))$, which is largest when $\hat{\mathbb{P}}(y_j = 1 | 1:i )$ is near 0.5 and shrinks as the probability approaches 0 or 1. Thus, after a sequence of contributions leading to a probability near 1, an additional positive contribution has minimal impact. In contrast, the same contribution would have a more substantial effect if added to a set of contributions ($1:(i-1)$) yielding a probability in the mid-range of the sigmoid function.  This is worth keeping in mind when visualizing contributions as a \textit{cumulative} time series that lands on the value of $\hat{\mathbb{P}}(y_j= 1 | 1:i )$, but depends at each $i$ on the sum of those that preceded it. One metric that we can monitor, which is not subject to this complication, is log-odds, because the mapping between contributions and the log-odds prediction is linear, i.e., 
\[
\log \left( \frac{\hat{\mathbb{P}}(y_j = 1 | 1:i )}{1 - \hat{\mathbb{P}}(y_j = 1 | 1:i )} \right) = \sum_{\iota=1}^{i} c_{j\iota}^{\text{log-odds}} \, .
\]
This way, contributions remain order-invariant and can provide complementary insights to visualizing  the marginal effect of $i$ on predicted probabilities. Nonetheless, in the empirical section, we report contributions to probabilities, as these are the units dominating recession risks discussions. We found that the aforementioned distortions were not overly obstructive in our applications. Moreover, marginal probability effects are not affected by the scale identification problems for coefficients in logistic (and probit) regressions.

\subsection{Comparison with Shapley Values}

As mentioned earlier, Shapley values are a natural candidate for interpretability tasks in machine learning, and \cite{ghorbani2019datashap} have proposed methods for using them to evaluate the importance of training data points. While Shapley values offer flexibility for complex models, they have certain limitations, particularly when the goal is to calculate contributions from training observations. First, the computational cost increases significantly with the size of the input data due to the need to calculate marginal contributions. Although approximations can reduce this cost, they often require restrictive or unconventional assumptions  \citep{rozemberczki2022shapley,kwon2023data}.  

Our approach, in contrast, directly utilizes model outputs without the need for resampling schemes. The computational complexity of our method is limited to evaluating various \(\lambda\) values for the ridge regression solution to closely match out-of-sample predictions—a task that can be completed in less than two seconds on a standard computer for the sample sizes we consider. In contrast, Shapley values or permutation-based importance measures for observations require re-estimating the model over a large number of runs. Furthermore, using Shapley values language, our dual approach also possesses the desirable "efficiency property", as the contributions sum exactly to the forecast being decomposed, by construction.

Beyond operational concerns, interpreting feature importance using Shapley values can be challenging. The axioms underpinning Shapley values are rooted in cooperative game theory, and their suitability for interpretability in machine learning models has seldom been questioned \citep{slack2020fooling,hooker2021unrestricted,verdinelli2023feature}. In our context, while leave-one-out methods would estimate the model excluding observation \(i\) to determine its marginal contribution in a model including the whole training set, Shapley values would calculate importance as the average of marginal contributions over all possible sub-training sets—including some less credible coalition featuring very few data points. Both concepts are relevant but distinct.  Still,  we prefer the simpler approach, where \(\boldsymbol{w}_j\) have a clear and intuitive interpretation as proximity coefficients.









 
\subsection{The Basic Toolbox}\label{sec:derivatives}

There are many ways to summarize the information in $\boldsymbol{w}_j$, and we summarize here the plotting schemes as well as the summary statistics being used in the empirical section.



 \vskip 0.15cm
{\noindent \sc \textbf{Time Series Plots}}.  A first step is to gaze at the time series of $\boldsymbol{w}_j$, either directly,  with a moving average, or through a normalized cumulative sum ending at one.  In our empirical exercises,  we opt for the moving average  and compare it throughout  to $ \frac{1}{N}\boldsymbol{1}$ which is the $\boldsymbol{w}_j$ implied by the unconditional mean forecast.   We also present contributions, $c_{ji} = w_{ji}y_i$, as moving averages and cumulative sums. The latter is particularly interesting,  as the constructed time series will land exactly on $\hat{y}_j$'s value. 
  

 
%

 \vskip 0.15cm

{\noindent \sc \textbf{Forecast Concentration}}. Concentration in distributions can be measured in several ways. We adopt a concentration ratio that reflects the proportion of the total sum of absolute weights contributed by the top $Q$\% of the weights:
\[
\text{\textbf{FC}}(\hat{y}_j) = \frac{\sum_{q=1}^{\lfloor Q \times \sfrac{N}{100} \rfloor} |w_{jq}|}{\sum_{q=1}^N |w_{jq}|} , 
\]
where $q$ denotes re-ordered weights from the largest absolute value ($q=1$) to the smallest ($q=N$). This approach echoes well-known measures used to characterize income and wealth inequality. The advantage of this metric lies in its interpretability; for instance, we can state that 50\% of \( \hat{y}_j \)'s forecast is driven by just 5\% of the observations, indicating significant concentration compared to the (equally weighted) unconditional mean, which would suggest a 5\% contribution. For algorithms such as random forest (RF), where \( \boldsymbol{w}_j \in \Delta^{N-1} \), alternative concentration metrics \textcolor{black}{could also be considered. One such example would be} the sum of squared weights, akin to a Herfindahl index that uses squared market shares to evaluate market concentration.

There is a clear link to forecast variance. The forecast with the smallest variance and highest bias is the unconditional mean. The \textbf{FC} metric indicates whether \( \hat{y}_j \) relies on a narrow or a broad set of observations. Naturally, one would feel more confident in a prediction if it draws from a diversified set of historical data rather than from just a few key events. On the opposite end of the spectrum,  a near-equally weighted portfolio suggests the forecast could be anybody's guess. 



 \vskip 0.15cm
{\noindent \sc \textbf{Forecast Short Position}}.  Short positions in forecasting can be risky, as they imply "borrowing" a mirrored $y_i$ (i.e., $-y_i$, not necessarily in the training data) to infer things about $y_j$ using symmetry assumptions.  For instance,  if CPI inflation went down massively in 2008 due to an oil shock that followed a significant contraction in economic activity,  then a rapid acceleration of economic activity can trigger an unrealistic prediction of extremely high inflation.  Such inferences can lead to important efficiency gains--symmetry implies using the same data point for more than one type of prediction--or significant misfires. For such reasons,  monitoring the extent of the forecast's "short position" appears worthwhile.  This can be done through
$$\text{\textbf{FSP}}(\hat{y}_j) = \sum_{i=1}^N I \left(w_{ji} < 0 \right) w_{ji} $$
and like $\textbf{FC}$, can be reported throughout the test set as a time series,  or is indicative in its own right for a single forecast to monitor to what extent mirrored training observations are utilized in a given prediction.



\vskip 0.15cm

{\noindent \sc \textbf{Forecast Leverage}}. The condition \(\sum_i^N w_{ji} = 1\) does not necessarily hold for a given \(j\) and certain classes of models. This introduces the concept of forecast leverage, defined as
\[
\text{\textbf{FL}}(\hat{y}_j) = \sum_{i=1}^N w_{ji},
\]
which is analogous to financial leverage in portfolio management. In finance, leverage allows an investor to amplify returns by borrowing, thereby increasing exposure beyond the initial capital (which would imply investing one unit and satisfying the constraint \(\sum_{i=1}^N w_{ji} = 1\)). Similarly, in the context of forecasting, if the sum of the weights \(w_{ji}\) is less than 1, the forecast is "underleveraged," meaning the influence of the in-sample observations on the forecasted value is diluted compared to a simple convex combination of in-sample observations. Conversely, if the sum exceeds 1, the forecast is "leveraged," amplifying the impact of the training data. This concept of forecast leverage provides a new perspective for understanding and interpreting the sensitivity and stability of model predictions, particularly when the model involves significant extrapolation (e.g., \(\sum_{i=1}^N w_{ji} > 1\)) or compression (\(0 \leq \sum_{i=1}^N w_{ji} < 1\)).


In linear models, forecast leverage is constrained by the structure of the projection matrix: in-sample, the sum of weights is exactly 1 for any prediction, a property inherent to models with an intercept. Out-of-sample, however, leverage can vary. For models with limited features capturing proximity—such as a small regression model or a neural network with few neurons in the final layer ($n_L$)—forecast leverage may deviate from 1, depending on the extent to which $X_j$ lies beyond the span of in-sample points $X_i$.

As feature dimensionality increases, forecast leverage tends to approach 1. This alignment results from a law-of-large-numbers effect, whereby out-of-sample regressors tend to fall closer, on average, to their in-sample counterparts, thereby stabilizing leverage near unity.

In ensemble models like random forest, forecast leverage is inherently fixed at one because predictions are weighted averages of in-sample observations. This fixed leverage, alongside the absence of short positions in RF, contributes to their robustness and sometimes overly conservative predictions \citep{MRFjae}.
%





 \vskip 0.15cm
{\noindent \sc \textbf{Forecast Turnover}}.  Turnover is a metric used to measure the frequency with which financial assets in a portfolio are bought and sold over a given period.  Tracking turnover is essential in evaluating portfolio performance because it provides insights into the trading activity within the portfolio. High turnover may indicate active management and could result in higher transaction costs, which can affect the overall returns.   

While such operational considerations do not map directly into macroeconomic forecasting,  one may still desire to monitor forecast turnover as a metric of plausibility.  Indeed,  while fast and large changes in $\boldsymbol{w}_j$ may sometimes reflect rapidly evolving conditions, they might, especially in calmer economic environments,  appear implausible.  As an example,  if current conditions are judged to most resemble the late 1970s,  it could be surprising that the following month or quarter should be heavily loading on 1997.  We can track this using 
$$\text{\textbf{FT}}(\hat{\boldsymbol{y}}) = \sum_{i=1}^N  \sum_{j=1}^J  | w_{ji} -w_{j-1,i} | $$ 
and compare it across models.  Unlike the portfolio evaluation metric from which it is inspired,  whether high $\text{\textbf{FT}}(\hat{\boldsymbol{y}})$ is a bad thing or a good one is context-dependent.

  \vskip 0.15cm
{\noindent \sc \textbf{Overall Historical Importance}}.  The main focus of the paper is that of decomposing a single out-of-sample prediction.  But,  for a given set of out-of-sample observations (OOS), one can calculate overall historical importance of a \textit{training} observation $i$ through

$$\text{\textbf{OHI}}(y_i) = \sum_{j \in \text{OOS}} |w_{ji}|,$$
which is analogous to an out-of-sample variable importance metric. \textcolor{black}{To evaluate the significance of various historical episodes, this measure becomes most informative when summed over specific intervals of the training sample. } 
 
\section{Empirical Application}\label{sec:emp}

In the empirical section, we provide new insights into interpreting ML forecasts in macroeconomics through our proposed approach.  For the regression case, we forecast key US macroeconomic variables—inflation, GDP growth, and unemployment changes. For classification, we predict US recessions.  We focus on selected OOS predictions in each application, for which we present weights and contributions of training observations underlying the prediction. Additionally, we assess these predictions using the statistical tools from Section \ref{sec:derivatives}.  Comprehensive forecasting performance evaluation and additional figures  are relegated to the appendix.

\subsection{Data and Models Setup}\label{sec:data}

The targets in our forecasting exercise include inflation (one step ahead), GDP growth (one step ahead), change in unemployment (one, two, four steps ahead), and recessions (three, twelve steps ahead) in the US, which we predict using a large-dimensional set of economic and financial aggregates. The data is taken from the popular FRED-QD/MD database of \cite{mccracken2020fred,mccrackenng} and transformed as suggested therein to induce stationarity. 
Our sample covers the period 1961Q2 to 2024Q1 at a quarterly frequency for inflation, GDP growth and unemployment, while we resort to the monthly frequency between 1961M4 to 2024M6 for our recession forecasting exercise.  We include four lags for quarterly data and 12 lags for the monthly setup. In case of our nonlinear ML models, we add moving averages of order 2, 4, and 8 for each variable, applying the Moving Average Rotation of X (MARX) transformation as recommended by \cite{MDTM}. Before entering the models, all series are standardized over the training sample to feature zero mean and unit variance.  

The versality of our proposition is illustrated by interpreting predictions obtained from various popular ML models. Following up on the discussion in Section \ref{sec:dual}, we include factor-augmented autoregression (FAAR), ridge regression (RR), kernel ridge regression (KRR), random forest (RF), Light Gradient Boosting Machine (LGB), and a deep neural network (NN). For our inflation application, we add the Hemisphere Neural Network (HNN) from \cite{HNN}, which models inflation using a nonlinear Phillips curve in the output layer.  This has been shown to improve predictive accuracy and interpretability for forecasting inflation, as it introduces some structure inspired by basic economic theory into otherwise highly flexible, nonlinear models \citep{HNN,GCFK}. 
We list model specifications and hyperparameters below.  Note that these settings are common to all applications.

\begin{enumerate}[leftmargin=5em, labelwidth=4em, labelsep=1em, align=left]
    \item[\texttt{ \fontfamily{phv}\selectfont \textbf{FAAR}:}] The FAAR includes four autoregressive lags along with two lags of the augmented factors. The latter are derived using standard principal components analysis, from which we extract four latent factors.
    \item[\texttt{ \fontfamily{phv}\selectfont \textbf{\phantom{FA}RR}:}] For RR, we cross-validate the penalty term $\lambda$ over a large set of possible values, allowing for heavy to minimal shrinkage. 
    \item[\texttt{ \fontfamily{phv}\selectfont \textbf{\phantom{F}KRR}:}] Hyperparameters for KRR are tuned via cross-validation. These include the choice of kernel (Laplace and Gaussian), bandwidth $\sigma$, and penalty term $\lambda$. We evaluate the model settings on 100 out-of-bag samples, using a subsampling rate of 80\% and shuffle blocks of eight quarters.  
    \item[\texttt{ \fontfamily{phv}\selectfont \textbf{\phantom{FA}RF}:}] Our RF features 500 trees, which ensures estimation stability for quarterly data. The \texttt{subsampling.rate} is set to 75\% and for the \texttt{block.size} we choose eight quarters. The \texttt{minimal.node.size}, which defines the smallest parent leaf size eligible for a new split, is set to 5. For the fraction of randomly selected predictors, \texttt{mtry}, we choose $\sfrac{1}{3}$, which is common practice to sufficiently diversify splitting candidates while keeping computational costs low.  
    \item[\texttt{ \fontfamily{phv}\selectfont \textbf{\phantom{F}LGB}:}] We also cross-validate the LGB. Here, the set of tuning parameters comprises the learning rate, maximum tree depth, the fractions of observations and features sampled to grow each tree. We take five out-of-bag samples with a sampling rate of 80\% in blocks of eight quarters. 
    \item[\texttt{ \fontfamily{phv}\selectfont \textbf{\phantom{FA}NN}:}] Our deep neural network is a standard feed-forward fully connected network featuring three hidden layers with 400 neurons each, and a linear output layer. For all nonlinear transformations, we choose the \textit{ReLU} activation function ($\operatorname{ReLU}(x)=\max \{0, x\}$). The maximum number of \texttt{epochs} is set to 100 and the \texttt{learning.rate} to 0.001. We introduce early stopping with a \texttt{subsampling.rate} of 85\% and a tolerance of 0.01 as well as dropout with a rate of 0.2. The \texttt{batch.size} is 32 periods and the number of bootstrap is 30. We use a mean squared error loss function and the standard Adam optimizer.
    \item[\texttt{ \fontfamily{phv}\selectfont \textbf{\phantom{F}HNN}:}] Following \cite{HNN}, the HNN consists of four hemispheres covering long-run and short-run expectations, output gap, and commodity prices. The variables feeding each hemisphere as well as all hyperparameter choices match those detailed in the original paper and are provided in Appendix \ref{sec:HNNprimer}. 
\end{enumerate}

This suite of models is fairly inclusive. FAAR is an easy-to-use, reliable macroeconomic forecasting model with a strong track record \citep{GCLSS2018}. However, its use of statistical factors to determine proximity in a reduced-dimension space introduces opacity to the forecasting process, making it a natural candidate for further interpretation through duality. RR is a standard linear ML model capable of handling high-dimensional inputs. Then, we have the three major families of modern nonlinear ML methods. First, kernel-based methods introduce nonlinearities using the well-known kernel trick. Second, \textcolor{black}{we use} two tree ensemble methods, renowned for their forecasting performance with tabular data \citep{grinsztajn2022tree,TBTP}. Third, two neural networks with reasonably straightforward architectures allow for nonlinearities through successive layers of nonlinear transformations. Together, these models constitute a comprehensive set of off-the-shelf ML techniques that have gained increasing popularity in recent years \citep{GCLSS2018}.

\subsection{Inflation for the Post-Pandemic Surge}\label{sec:infl}

Our first application examines inflation predictions for the post-pandemic era, using a hold-out sample that ranges from 2020Q1 to 2024Q1. We select three key dates that cover important phases of the recent inflation cycle: 2020Q3 (initial Covid-19 shock), 2021Q2 (the awakening), and 2022Q2 (the peak). 
In our discussion, we focus on LGB, as our best performing tree-based model, RR, as the best performing linear model, NN, HNN, and KRR. 

\begin{figure}
\caption{\normalsize{Dual Interpretation of Inflation Predictions ($h=1$)}} \label{fig:infl}
\centering
\vspace*{-0.5em}
      \includegraphics[width=\textwidth, trim = 0mm 0mm 3mm 0mm, clip]{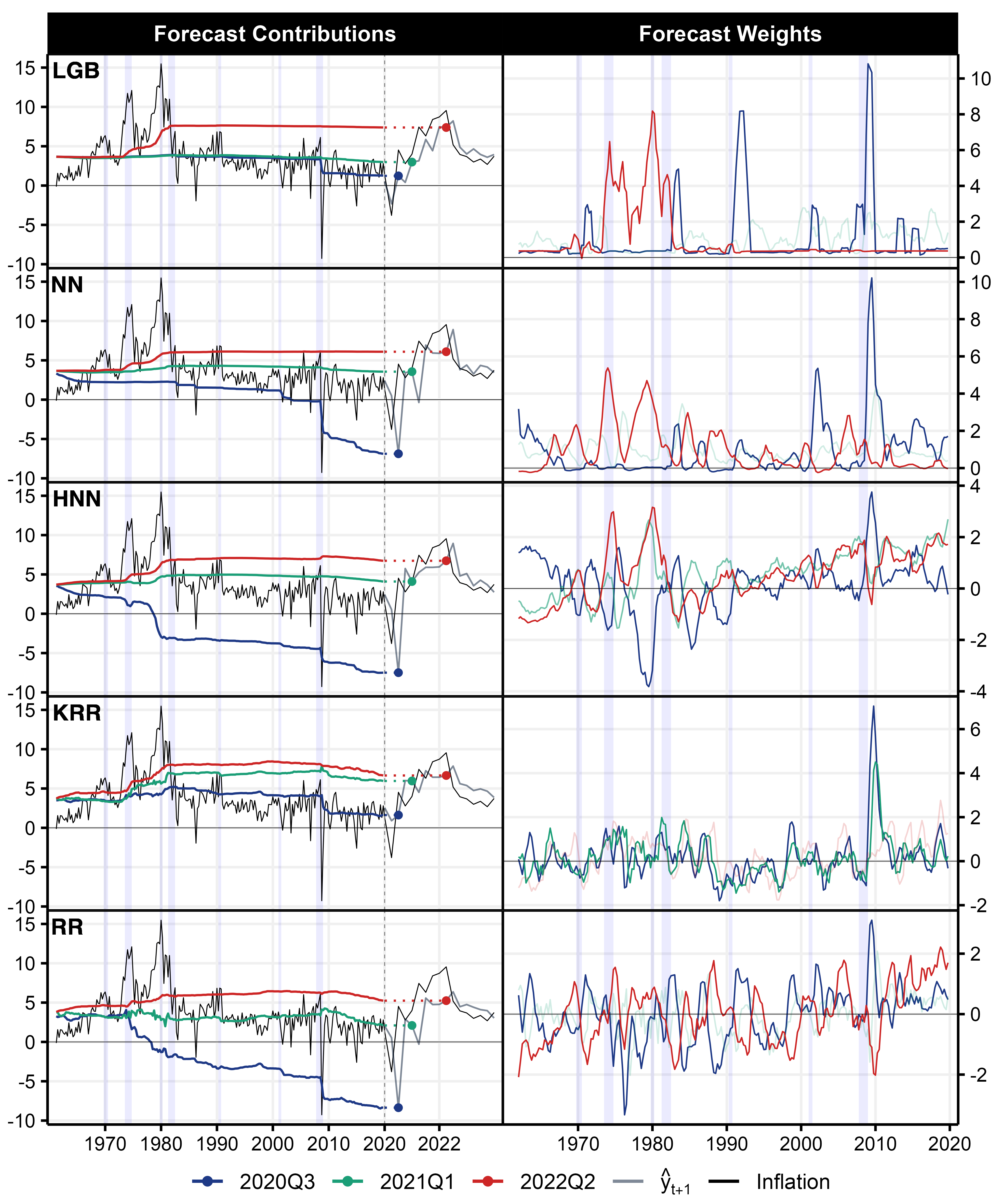}

 \begin{threeparttable}
    \centering
    \vspace*{-1.5em}
    \begin{minipage}{\textwidth}
      \begin{tablenotes}[para,flushleft]
    \setlength{\lineskip}{0.2ex}
    \footscript 
  {\textit{Notes}: The figure presents results from predicting CPI inflation one step ahead. The \textbf{left panels} present the \textit{cumulative} sum of forecast contributions $c_{ji}$} over the training sample (1961Q2 to 2019Q4), which collectively sum to the final predicted value $\hat{y}_j$ shown as dots. We initialize $c_{j0}$ at the unconditional average of the sample and present $c_{ji}$ as deviations from this average. The holdout sample ranges from 2020Q1 to 2024Q1, indicated by the dashed line. The \textbf{right panels} show forecast weights $w_j$ as a moving average of four quarters scaled by the mean of absolute weights. Lavender shading corresponds to NBER recessions. We selectively fade certain lines to enhance visibility, especially for models with high turnover.
    \end{tablenotes}
  \end{minipage}
  \end{threeparttable}
\end{figure}

Figure \ref{fig:infl} illustrates the cumulative sum of contributions over time ($c_{ji}$ for $j \in \{\text{2020Q3}, \text{2021Q2}, \text{2022Q2}\}$ and $i$ from $\text{1961Q2}$ to $\text{2019Q4}$), converging to the final predicted value $\hat{y}_{j}$. To ensure comparability across models and target variables, we initialize all $c_{j0}$ at the unconditional mean of the training sample and present $c_{ji}$ as deviations from this average. This adjustment is particularly important for tree-based ensembles applied to predominantly positive target variables, where the raw cumulative sum naturally starts at zero and rises to a positive value. Consequently, unadjusted time series plots would exhibit an upward trend that trivially reflects the accumulation of mostly positive observations. Therefore, the deviations from this mechanical trend— expressing contributions as deviations from an egalitarian proximity forecast—elicit the pertinent information and are universally comparable.



In the right panels, we plot the time series of weights, which, as discussed in Section \ref{sec:dual}, can be directly interpreted as proximity scores. For ease of visualization, especially when plotting many series at once, we use a moving average of four quarters. To provide an alternative view of forecast contributions, we also present $c_{ji}$ as a four quarters moving average in Figure \ref{fig:contribma4_infl} in the appendix. Table \ref{tab:metrics_infl} summarizes the remaining statistics from Section \ref{sec:derivatives}, including forecast concentration, short position, leverage, and turnover. Results on overall historical importance (Figure \ref{fig:ohi}) and the models' forecasting performance (Table \ref{tab:rmse}) can be found in Appendix \ref{sec:addresults}.


\vskip 0.15cm
{\noindent \sc \textbf{Initial Covid-19 Shock (\textcolor{darkblue2}{2020Q3})}}.   Unsurprisingly, the initial Covid-19 shock in 2020Q3 caused many models to produce poor inflation forecasts. The unprecedented shutdown of the economy made projecting the future trajectory of inflation exceptionally challenging.  While one can speculate about the implausible associations with past historical events that a model might use, interpretation via duality can provide a crisp quantitative answer as to where failures are rooted.  For all ML models, we find that their efforts to leverage information from the past following the original Covid-19 shock result in excessive weighting towards the GFC.  However, unlike during the GFC, the brief but pronounced stock market crash and sharp decline in real activity following the Covid-19 outbreak did not lead to a comparable drop in the inflation rate \citep{meyer2022covid,bobeica2023covid}. While the risk of drawing parallels to the GFC was obvious to any analyst with access to external, non-model information in 2020, it is much less so for an algorithm relying solely on traditional macroeconomic data. Still, it is interesting to analyze how ML tools perceived the initial Covid-19 shock, and "tried their best" at finding similarities with the past \textcolor{black}{to come up with a number.} 

Going into more details, we can identify both similarities and differences between the ML models' dual interpretation for this particular prediction. First, both LGB and NN see parallels with other past crises, such as those in the 1980s, 1990s, and the early 2000s, which complement the strong emphasis on the GFC (see right panels of Figure \ref{fig:infl}). Noteworthy, we find NN assigning considerably high weight on both the 2001 recession and the GFC, which leads to strong negative contributions (see left panels \textcolor{black}{and, alternatively, Figure \ref{fig:contribma4_infl} in the appendix}). Two quarters stand out: 1) 2001Q4 with a raw weight of nearly 10\%, a period following the 9/11 attacks and within the aftermath of the Dotcom bubble, and 2) 2009Q1 with over 20\% in raw weights, when stock markets hit record lows in March and unemployment surged to peak levels.  High concentration and high leverage found for NN (see Table \ref{tab:metrics_infl}) amplifies the influence of those training observations  and, in this case, substantially worsens the final prediction.  LGB is also heavily concentrated, with the top 10 training observations comprising nearly half of the portfolio. However, only a small fraction of these observations correspond to highly negative values of $\boldsymbol{y}$, which mitigates the loss in LGB's forecast for 2020Q3. 





HNN and RR not only perceive similarities with the GFC but also recognize substantial dissimilarities with the second inflation surge of the 1970s. This is evident from the significant negative weights both models assign to this period, reflected by the negative values in terms of forecast short position (see Table \ref{tab:metrics_infl}).  This proves to be a remarkable mistake.  As shown in the left panels \textcolor{black}{and Figure \ref{fig:contribma4_infl} in the appendix}, both the 1970s and the GFC contribute negatively to the models' forecasts in 2020Q3. Despite relatively low forecast leverage and concentration, these shortcomings are sufficient to deliver a massive forecast error.  An analyst presented with the proximity information ex-ante could manually remove the aforementioned implausible contributions and, in both cases, bring the forecasts out of deflation territory.

For KRR, as with the other models, we observe significant weight assigned to the GFC, leading to a negative contribution to the forecast. However, its effect is less pronounced and more comparable to that of LGB, which helps keep the forecast modestly positive. The forecast concentration for the 2020Q3 prediction is mid-range among competitors but relatively high compared to the model's other out-of-sample predictions. The forecast short position indicates balancing forces between upward and downward movements of inflation.

\begin{table}[t]
\vspace{1em}
  \footnotesize
  \centering
  \begin{threeparttable}
  \caption{\normalsize {Forecast Statistics for Inflation ($h=1$)} \label{tab:metrics_infl}
    \vspace{-0.3cm}}
    \setlength{\tabcolsep}{0.6em} 
    \setlength\extrarowheight{2.9pt}

    \begin{tabular}{l| rrr|rrr|rrr|r}
    \toprule \toprule
    \addlinespace[2pt]
    \multicolumn{1}{l|}{} & \multicolumn{3}{c|}{Concentration}  & \multicolumn{3}{c|}{Leverage} & \multicolumn{3}{c|}{Short Position} & \multicolumn{1}{c}{Turnover}  \\ 
    \cmidrule(lr){2-4} \cmidrule(lr){5-7} \cmidrule(lr){8-10} \cmidrule(lr){11-11}
    \multicolumn{1}{r|}{} & \multicolumn{1}{c}{2020Q3}  & \multicolumn{1}{c}{2021Q1} & \multicolumn{1}{c|}{2022Q2} & \multicolumn{1}{c}{2020Q3}  & \multicolumn{1}{c}{2021Q1} & \multicolumn{1}{c|}{2022Q2}  & \multicolumn{1}{c}{2020Q3}  & \multicolumn{1}{c}{2021Q1} & \multicolumn{1}{c|}{2022Q2} & \multicolumn{1}{c}{Overall} \\ 
    \midrule
  LGB & 0.49 & 0.18 & 0.36 & 1.00 & 1.00 & 1.00 & 0 & 0 & 0 & 12.76 \\ 
  RF & 0.51 & 0.15 & 0.33 & 1.00 & 1.00 & 1.00 & 0 & 0 & 0 & 10.15 \\ 
  NN & 0.35 & 0.19 & 0.21 & 3.18 & 1.36 & 0.93 & -0.13 & -0.01 & -0.02 & 15.71 \\ 
  HNN & 0.18 & 0.13 & 0.14 & 0.43 & 0.78 & 0.85 & -2.41 & -0.34 & -0.50 & 26.97 \\ 
  KRR & 0.22 & 0.19 & 0.15 & 0.99 & 0.99 & 1.00 & -2.42 & -2.61 & -1.97 & 89.55 \\ 
  RR & 0.16 & 0.17 & 0.13 & 1.00 & 1.00 & 1.00 & -5.27 & -4.88 & -1.68 & 97.79 \\ 
  FAAR & 0.16 & 0.18 & 0.16 & 1.00 & 1.00 & 1.00 & -18.78 & -5.16 & -1.14 & 264.86 \\ 
   \bottomrule \bottomrule 
\end{tabular}
\begin{tablenotes}[para,flushleft]
  \footscript 
    \textit{Notes}: The table summarizes forecast metrics as discussed in Section \ref{sec:derivatives}. Forecast concentration shows the proportion of total weights attributed to the top 5\% of the weights ($Q=5$).
  \end{tablenotes}
\end{threeparttable}
\end{table}

\vskip 0.15cm
{\noindent \sc \textbf{The Awakening (\textcolor{darkgreen2}{2021Q1})}}. For the first stages of the inflation surge in 2021Q1 many models underappreciate the mounting upward pressures and produce muted forecasts.  Most of them (LGB,  NN,   RF,  FAAR) return approximately the unconditional mean (3.6\%, based on the 1961Q2-2019Q4 sample).  This is above the Federal Reserve's target range and the average of the last two decades, but still below the realized value of 4\%. Weights are rather equally distributed over the training observations, reflected by low forecast concentration across models.  Forecast leverages are close to one for most models, except for our neural network specifications, and forecast short positions show some balancing acts between positive and negative pressures.   In sum, most models do not know where to look for and discriminate for relevant information---it does not yet look like 1970s, nor does it seem like a continuation of the 2010s low and stable inflation era.  This results in the conservative equally weighting of the training sample.  Despite lacking luster, this proves more fruitful a strategy than that of RR, which finds strong resemblance to the 2010s and report a forecast close to the 2\% target. 



HNN and KRR stand out by predicting elevated inflation levels for 2021Q1. An analyst monitoring a panel of ML-based forecasting models in late 2020 would surely wonder what these two models see that the others do not, and how proximity information can be used to weigh in on the debate over whether inflation is transitory or persistent.  We find that HNN and KRR share the common trait of being the only models observing similarities with the high-inflation period of the 1970s. 

HNN's prediction receives positive contributions from the second run-up of inflation in the late 1970s, with the highest (raw) weight in 1979Q2 at the beginning of the second oil crisis. It also features a lasting proximity to recent inflation conditions, which mildly dampens the forecast by integrating post-great recession data onward. Therefore, at this juncture, HNN is still ambivalent, seeing similarity to only one of the two 1970s inflation surges while still featuring a significant share of Great Moderation data points. 

KRR is more decisive, but this leads to a positive prediction error.  Indeed, its forecast is pushed up by both the first and second 1970s surges in inflation. Additionally, KRR assigns positive weight to the final quarters of the GFC (2009Q3 to 2010Q1), exerting slight downward pressure on the prediction, which helps prevent severe overshooting of the actual value.

\vskip 0.15cm
{\noindent \sc \textbf{The Peak (\textcolor{darkred2}{2022Q2})}}.  As inflation reached its peak in 2022Q2, all models recognized the similarities with the high-inflation periods of the 1970s. Consequently, we observe noteworthy parallels being leveraged by the machine learning models.  This is not surprising, as the supply-driven sources of the recent inflation surge and the energy price jumps resemble the oil shocks of 1973 and 1979–80. Moreover, monetary policy remained accommodative during the inflation run-ups in these historical episodes \citep{gagliardone2023oil,ball2022understanding}.

More specifically, LGB and NN assign high weight to both inflationary episodes in the 1970s, with the second surge contributing particularly strongly to the forecast. Raw weights peak in 1973Q3, marking the beginning of the first oil crisis, and in 1978Q4, when the Iranian Revolution triggered the second dramatic increase in oil prices. For both models, we observe an uptick in forecast concentration compared to 2021Q1. Additionally, forecast leverage for NN decreased, indicating some compression in the impact of training data.

At the peak,  HNN and KRR forecasts are now driven by \textit{both} inflation run-ups of the 1970s. The highest weights for both instances are observed a few quarters later than in LGB and NN. For HNN, the peak weights are found in 1974Q2 for the first oil crisis and in 1980Q1 for the second. For KRR, they occur in 1975Q2 and 1980Q3. In addition, both models assign positive weight to more recent periods in the training set between the GFC and Covid-19, which slightly dampers the forecast. Lastly, RR does not associate 2022Q2 conditions with those prevailing in the 1970s and therefore, for the most part, fails to capture the post-2020 inflation surge.

\begin{figure}
\caption{\normalsize{An Alternative View: Moving Average Forecast Contributions for Inflation ($h=1$)}} \label{fig:contribma4_main}
\centering
\vspace*{-0.5em}
      \includegraphics[width=\textwidth, trim = 0mm 0mm 3mm 0mm, clip]{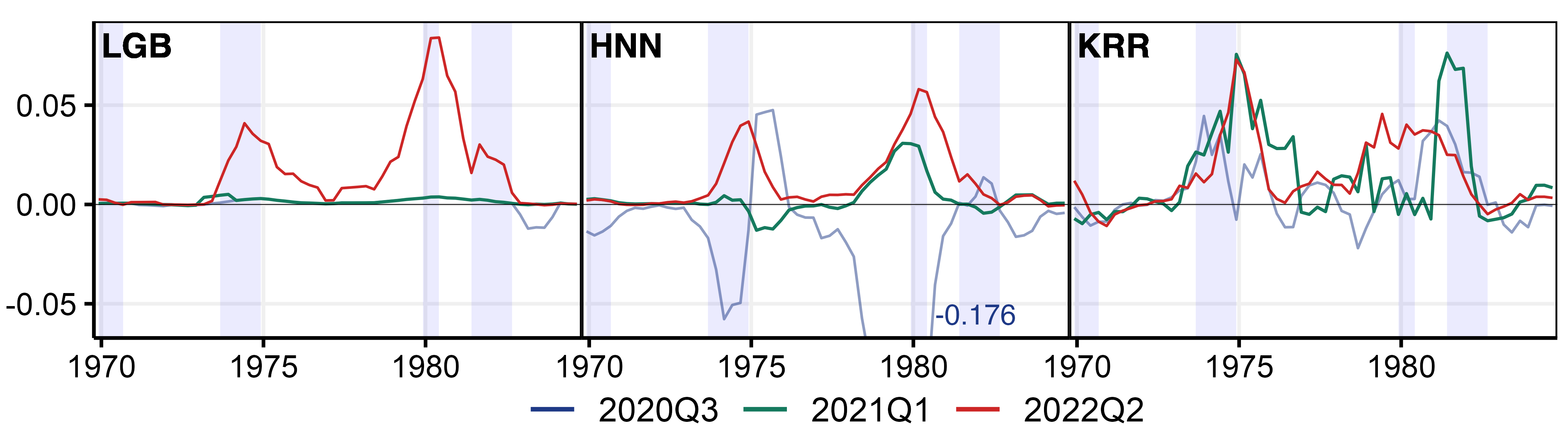}

 \begin{threeparttable}
    \centering
    \vspace*{-1.5em}
    \begin{minipage}{\textwidth}
      \begin{tablenotes}[para,flushleft]
    \setlength{\lineskip}{0.2ex}
    \footscript 
  {\textit{Notes}: The figure presents $c_{ji}$ as a moving average of four quarters. Lavender shading corresponds to NBER recessions.}
    \end{tablenotes}
  \end{minipage}
  \end{threeparttable}
      \vspace*{-1.5em}
\end{figure}

\vskip 0.15cm
{\noindent \sc \textbf{An Alternative Representation}}.   Figure \ref{fig:contribma4_main} focuses on the 1970-1990 period for three models displaying distinct behaviors. We use a moving average view of contributions, which, unlike the right panel of Figure \ref{fig:infl}, highlights instances where \textit{both} the proximity score and the training target observation are elevated. Thus, this representation has the potential to be sparser. First, we observe clear alignment in the origins of contributions for 2022Q2. Second, we see the sources of heterogeneity behind the 2021Q1 forecasts:  LGB shows no contributions from either inflation spike, HNN captures gentle contributions from the second surge, and KRR collects from both. 
 While KRR's weights for the onset of the surge appear lower in Figure \ref{fig:infl}, the alternative representation makes these instances stand out. Thus, we emphasize that no single reporting method dominates for conveying the dual solution’s information; it depends on the application.

\begin{figure}
\caption{\normalsize{Dual Interpretation of Additional Inflation Predictions ($h=1$)}} \label{fig:inflspecial}
\centering
\vspace*{-0.5em}
      \includegraphics[width=\textwidth, trim = 0mm 0mm 3mm 0mm, clip]{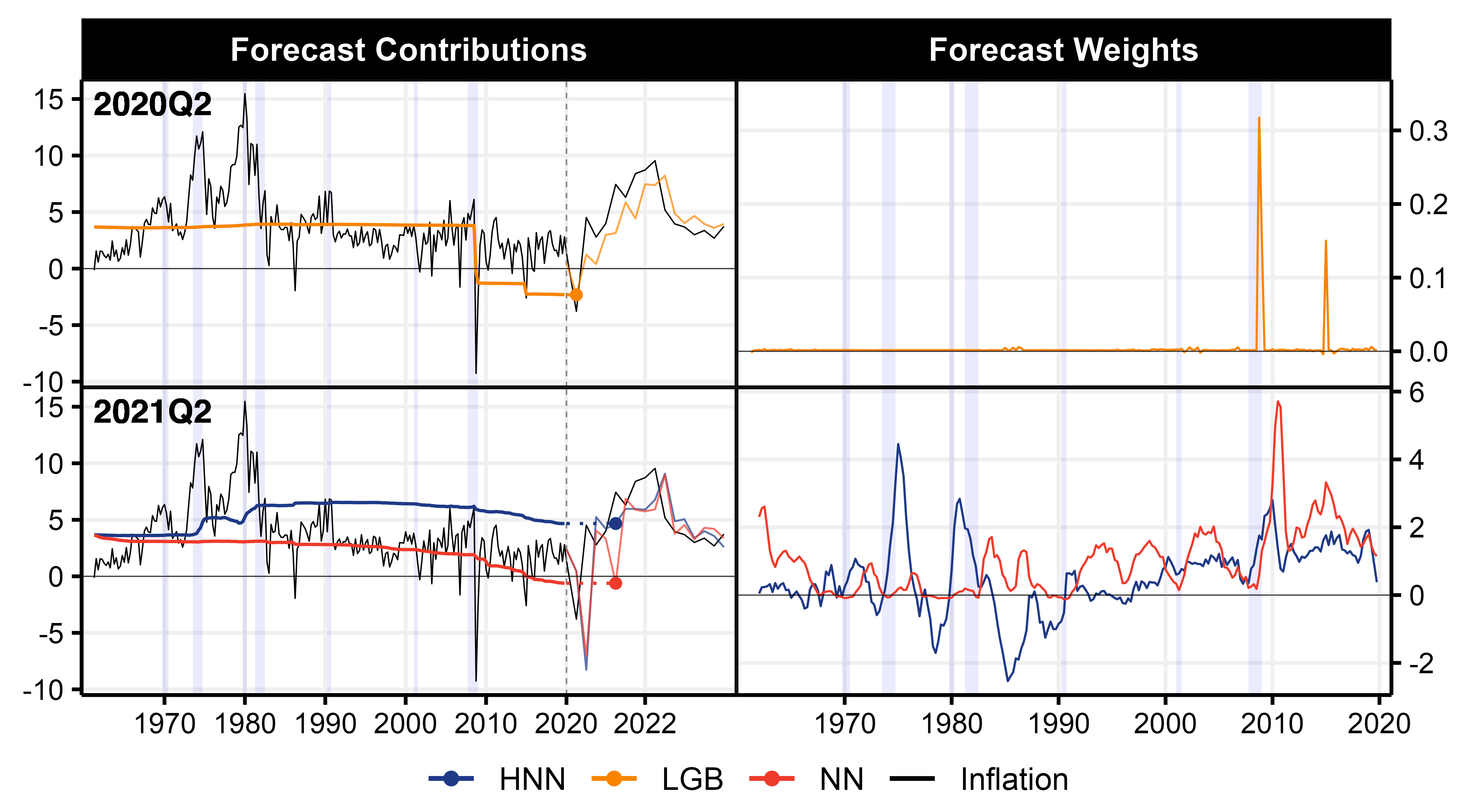}

 \begin{threeparttable}
    \centering
    \vspace*{-1.5em}
    \begin{minipage}{\textwidth}
      \begin{tablenotes}[para,flushleft]
    \setlength{\lineskip}{0.2ex}
    \footscript 
  {\textit{Notes}: For more details we refer to Figure \ref{fig:infl}. For illustrative purposes, we present raw weights in the upper right panel, while we take $w_j$ as a moving average of four quarters scaled by the mean of absolute weights in the lower right panel. Note that in the left panels colored lines during the holdout sample (after the dashed line) refer to the final predictions for each model and horizon.}
    \end{tablenotes}
  \end{minipage}
  \end{threeparttable}
\end{figure}


\vskip 0.15cm
{\noindent \sc \textbf{Other Intriguing Cases}}.  Two additional predictions that merit attention are those from LGB for 2020Q2 and NN for 2021Q2.  We present them in Figure \ref{fig:inflspecial}. LGB surprises with its highly accurate downward prediction for 2020Q2, even before Covid-19's impact on the economy is fully measured by quarterly data (only the end of March 2020 is included in Q1). Two periods stand out that lead to this outcome: the GFC (2008Q4 and 2009Q1) and the sovereign debt crisis in Europe (2015Q1). LGB leverages key commonalities these periods share, such as high uncertainty and recessionary tendencies—including a flattening of the yield curve, highly volatile stock markets, and low to negative growth—all implying downward pressures on inflation.

The second example, NN for 2021Q2, is surprising for being far off the mark.  NN predicts a sharp decline in inflation at a time when it is already on the rise. This is all the more noticeable given that NN's predictions for 2021Q1 and 2021Q3 are more than decent. In contrast, HNN is fairly consistent from one quarter to the next, and accurately predicts 5\% for 2021Q2.

Therefore, it begs the question of whether the dual decomposition could help discipline NN's forecasts ex-ante when such oddities arise.  NN’s misinterpretation of similarities has several roots.    It almost entirely ignores high-inflation periods like the 1970s and 1980s, instead drawing parallels to downward-trending inflation periods, including the mid-1980s, turbulent times like 2009Q1 and 2015Q1, and the expansion phase between the GFC and Covid-19.  These mistakes compound into a staggering -7.5\% forecast error.  We also see a gradually increasing weighting of recent observations in HNN, but it is completely offset by allocating sizable attention to the two inflation spikes of the 1970s—episodes that are entirely muted in NN’s 2021Q2 prediction.

\subsection{GDP Growth During the Great Recession}\label{sec:gdp}

\begin{figure}
\caption{\normalsize{Dual Interpretation of GDP Growth Predictions}} \label{fig:gdp}
\centering
\vspace*{-0.5em}
      \includegraphics[width=\textwidth, trim = 0mm 0mm 3mm 0mm, clip]{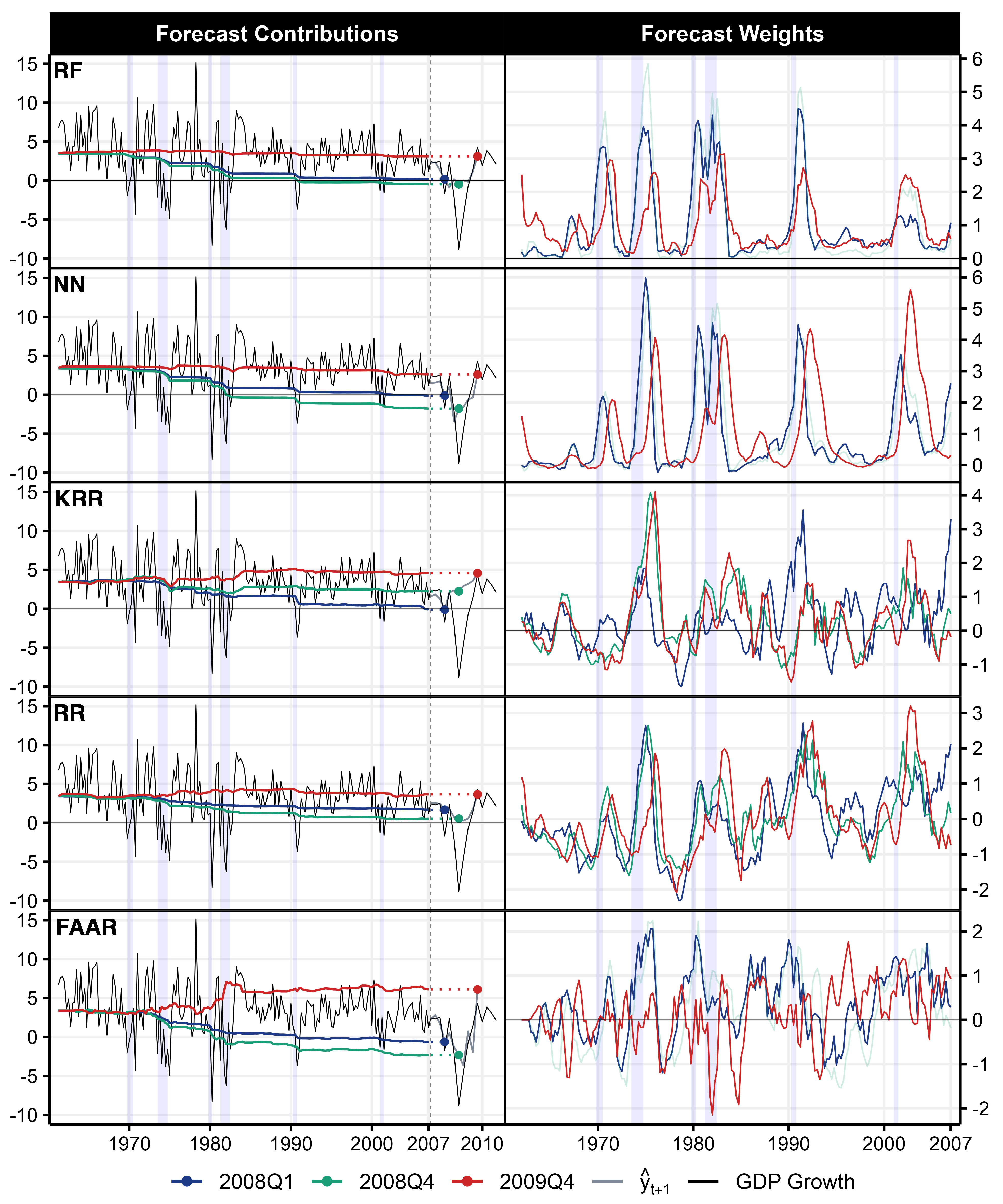}

 \begin{threeparttable}
    \centering
    \vspace*{-1.5em}
    \begin{minipage}{\textwidth}
      \begin{tablenotes}[para,flushleft]
    \setlength{\lineskip}{0.2ex}
    \footscript 
  {\textit{Notes}: The figure presents results from predicting GDP growth one step ahead. The \textbf{left panels} present the \textit{cumulative} sum of forecast contributions $c_{ji}$ over the training sample (1961Q2 to 2007Q1), which collectively sum to the final predicted value $\hat{y}_j$ shown as dots. We initialize $c_{j0}$ at the unconditional average of the sample and present $c_{ji}$ as deviations from this average. The holdout sample ranges from 2007Q2 to 2009Q4, indicated by the dashed line. The \textbf{right panels} show forecast weights $w_j$ as a moving average of four quarters scaled by the mean of absolute weights. Lavender shading corresponds to NBER recessions. We selectively fade certain lines to enhance visibility, especially for models with high turnover.}
    \end{tablenotes}
  \end{minipage}
  \end{threeparttable}
\end{figure}

Our second experiment is GDP growth during the GFC. We choose an out-of-sample period running from 2007Q2 to 2009Q4 and a one step ahead forecast horizon ($h=1$). 
The three quarters we will  focus on are 2008Q1 and 2008Q4 (the recession), and 2009Q4 (recovery). 
This selection allows to capture three distinct phases of the GFC: a relatively mild first half, followed by a sharp deepening of the recession in late 2008, and finally, the recovery at the end of 2009.  Results are presented in Figure \ref{fig:gdp} with contributions to the final prediction in the left panels and forecast weights in the right panels. \textcolor{black}{Figure \ref{fig:contribma4_gdp} (Appendix) presents contributions as moving average of four quarters.} In Tables \ref{tab:metrics_gdp} and  \ref{tab:rmse} (Appendix), we report the forecast summary statistics and overall forecast performance metrics, respectively.

\vskip 0.15cm
{\noindent \sc \textbf{The Recession (\textcolor{darkblue2}{2008Q1}, \textcolor{darkgreen2}{2008Q4})}}. During the first half of 2008, the contraction caused by emerging financial disruptions was comparably mild. GDP growth first turned negative in 2008Q1, which most models capture remarkably well with predictions of slightly negative or zero growth. The only exception is RR, which predicts a low but positive growth rate. The recession then aggravated with the collapse of Lehman Brothers in September 2008, triggering one of the sharpest stock market crashes in recent history the following month. Though recognizing signs of growing uncertainty, the severity of the contraction in 2008Q4 catches the models by surprise. The majority fails to anticipate the extent of the downturn, with only NN and FAAR suggesting noticeably negative growth rates. 

The final predictions, shown in the left panels of Figure \ref{fig:gdp}, reveal that NN and FAAR are the models most confident about an impending recession in early 2008--despite the absence of any negative GDP growth readings at that point. They are also the ones anticipating a deeper contraction in 2008Q4. In particular, NN's accuracy hinges on drawing similarities with every major recession in the training sample and leveraging conditions that recur across periods of economic disruptions, such as oil price run-ups \citep{kilian2008,hamilton2011}, shifts in consumer confidence \citep{blanchard1993}, and tight credit markets \citep{brunnermeier2009}.  

Relatively similar weighting schemes are observed for RF and RR, but with overall downscaled weights \textit{and} little to no weight at all on the 2001 recession. This leads to more subdued predictions for both the initial downturn as well as the trough. 

Notably, FAAR also finds key differences with expansionary periods, as is evident from the right panels in Figure \ref{fig:gdp} and the significant short position presented in Table \ref{tab:metrics_gdp} (Appendix). Both the Great Moderation and the quarters between the oil crises add further downward pressure to its final prediction. 
This stands in stark contrast to, e.g., the 2008Q4 prediction of KRR, which relies on similarities with the Great Moderation, pushing its forecast back up into misdirected positive territory. 

\vskip 0.15cm
{\noindent \sc \textbf{Recovery (\textcolor{darkred2}{2009Q4})}}. The deep recession at the end of 2008 was followed by a slow and sluggish recovery. All models effectively predict the rebound in 2009Q4. While FAAR is the only model that overshoots the target, we generally find high yet accurate final predictions.

Starting with the forecast weights presented in the right panels of Figure \ref{fig:gdp}, we observe that all models tend to assign high weight to the quarters marking the onset of recovery phases after each prolonged recession. Compared to contractionary forecasts, which we show are often characterized by a distinct spike that emphasizes a single quarter, the weights for the rebound forecasts are more right-skewed. While RF and NN identify clear similarities across each expansionary episode, KRR places greater emphasis on the periods following the 1970s recession. 

With this knowledge in mind, shifting our focus to the left panels of Figure \ref{fig:gdp} reveals an important insight: a flat line for contributions, which keeps the forecast aligned with the unconditional mean, does not necessarily imply that the models fail to recognize similarities or dissimilarities with past periods. In the case of the 2009Q4 prediction, the models capture the relevant expansionary phases in the business cycle, but during these periods GDP growth was evolving near its unconditional mean. This results in the observed flat line. 

FAAR's overly optimistic prediction derives most of its contributions from the 1970s and 1980s, especially dissimilarities found with the twin recessions.  This historical period raises FAAR’s forecast from the 3\% range to above 6\%. The contribution of the 1981-1982 recession particularly stands out in Figure \ref{fig:contribma4_gdp} (Appendix).   The importance of the short positions increases significantly (Table \ref{tab:metrics_gdp}, Appendix), misdirecting the forecast towards that of an overly optimistic rebound. Given that the severe recessions in the 1970s and 1980s transitioned into high output growth above long-run trends \citep{dominguez2013,fernald2018}, the overstated optimism in FAAR's recovery forecast is less surprising.  

\subsection{Unemployment at Different Horizons for the Great Recession Peak}\label{sec:unemp}

\begin{figure}
\caption{\normalsize{Dual Interpretation of Unemployment Predictions (2009Q1)}} \label{fig:unemp}
\centering
\vspace*{-0.5em}
      \includegraphics[width=\textwidth, trim = 0mm 0mm 3mm 0mm, clip]{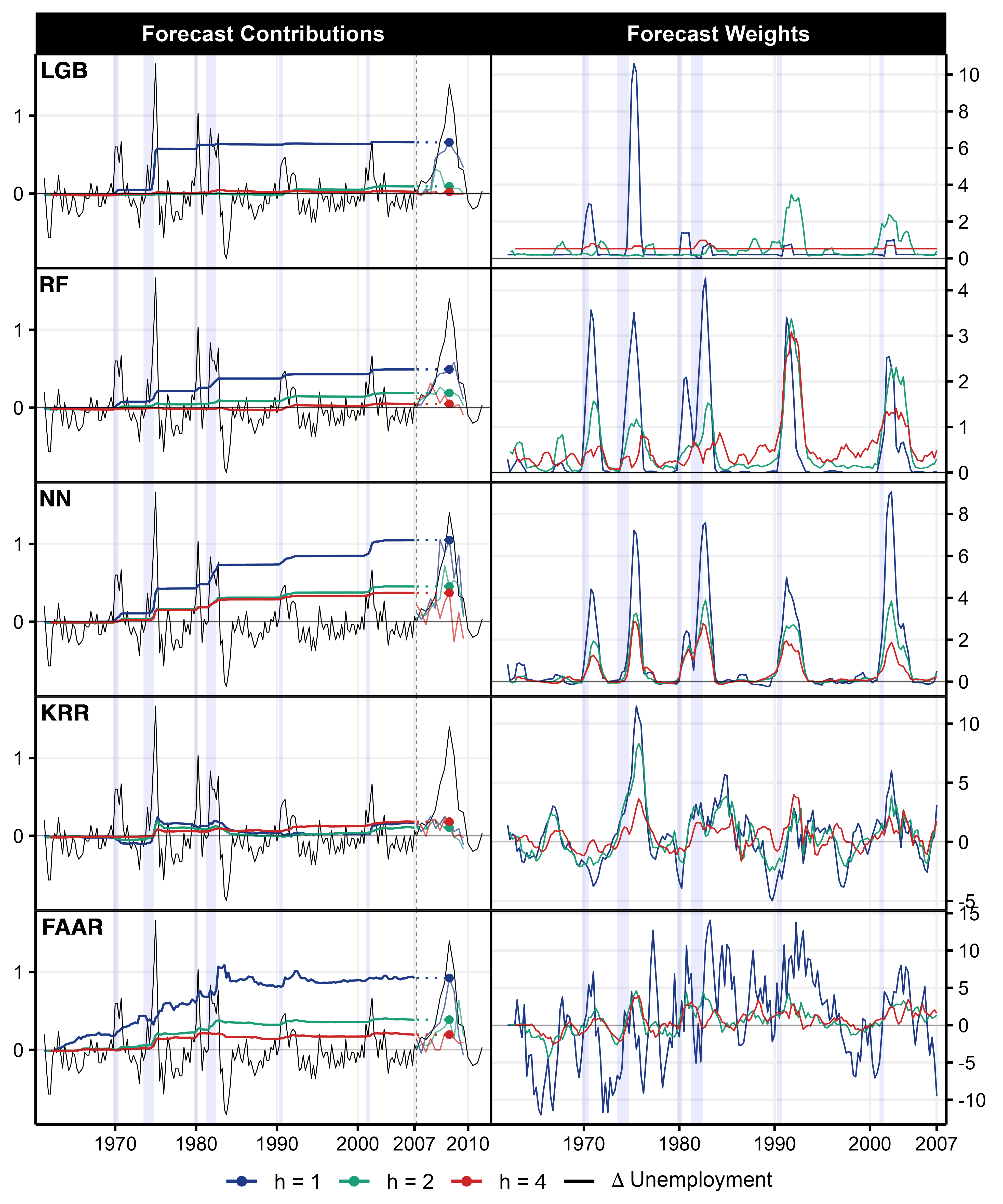}

 \begin{threeparttable}
    \centering
    \vspace*{-1.5em}
    \begin{minipage}{\textwidth}
      \begin{tablenotes}[para,flushleft]
    \setlength{\lineskip}{0.2ex}
    \footscript 
  {\textit{Notes}: The figure presents results from predicting the change in the unemployment rate one, two, and four steps ahead. The \textbf{left panels} present the \textit{cumulative} sum of forecast contributions $c_{ji}$ over the training sample (1961Q2 to 2007Q1), which collectively sum to the final predicted value $\hat{y}_j$ shown as dots. We initialize $c_{j0}$ at the unconditional average of the sample and present $c_{ji}$ as deviations from this average. The holdout sample ranges from 2007Q2 to 2009Q4, indicated by the dashed line. Note that colored lines during the holdout sample (after the dashed line) refer to the final predictions for each model and horizon. The \textbf{right panels} show forecast weights $w_j$ as a moving average of four quarters. Lavender shading corresponds to NBER recessions.}
    \end{tablenotes}
  \end{minipage}
  \end{threeparttable}
\end{figure}

In this subsection, we investigate predictions for the change in unemployment at various horizons ($h=\{1,2,4\}$) during the GFC. We forecast out-of-sample the period from 2007Q2 to 2009Q4 and focus on quarter 2009Q1, which marks the peak in the change of the unemployment rate during this crisis episode. We present results on forecast contributions and weights in Figure \ref{fig:unemp}, and \textcolor{black}{show moving averages of contributions in Figure \ref{fig:contribma4_unemp} (Appendix)}. Our usual set of forecast metrics can be found in Tables \ref{tab:metrics_unemp} and \ref{tab:rmse} in the appendix. 

\vskip 0.15cm
{\noindent \sc \textbf{The Peak for \textcolor{darkblue2}{$h=1$}}}. By the end of the first quarter in 2009, the unemployment rate surged to its peak, reflecting the severe economic fallout from the GFC. Most models in our set capture these dynamics well in their one step ahead predictions, with NN yielding the most accurate forecast, followed by FAAR and the tree-based specifications. KRR, on the other hand, diverges from the pack by suggesting only muted upward pressures. Overall, we find high forecast concentration for this particular prediction and high turnover compared to other forecast horizons.


NN, as our leading model, assigns high weight to each major recession in the training set, resulting in step function-like contributions. 
It receives the largest contributions from both oil crises (1975Q1 and 1982Q2), which rank among the longest and deepest post-war recessions preceding the GFC, as well as the early 2000s recession (2001Q4), which, though relatively mild and brief, resulted in persistent unemployment \citep[also referred to as "jobless recovery", see, ][]{kliesen2003,sinai2010}. As we find for GDP growth in Section \ref{sec:gdp}, NN proximity scores very much aligns with that of RF except for the 2001 recession. Again, RF's more muted similarity assessment leads to a more conservative forecast whereas NN benefits from the additional push.

While NN and FAAR report similar forecasts, the underlying inferences differ. Indeed,  FAAR  puts a distinctive accent on the pre-Great Moderation part of the sample, drawing on both similarities and dissimilarities. We observe substantial contributions to the forecast from dissimilarities to the 1960s and early 1970s--periods characterized by low unemployment rates--along with positive weights on both oil crises and the 1990s recession. Figure \ref{fig:contribma4_unemp} in the appendix, which plots moving averages of contributions, clarifies the positive impact of dissimilarities with expansionary episodes.


Of all models considered, LGB stands out for its sparsity. For a forecaster using LGB,  the highly concentrated nature of the prediction (66\% of the prediction is driven by the top 5\% of weights in Table \ref{tab:metrics_unemp}) would warrant further scrutiny. We find that LGB prediction for 2009Q1 ($h=1$) relies primarily on the first oil crisis of the 1970s. In particular, the model picks out 1975Q1, a critical quarter for several reasons. First, it directly precedes the peak in the unemployment rate observed during the first oil crisis (in May 1975). Second, it follows the oil price peak observed in 1974, drawing a parallel to the surge in 2008Q3. Oil price surges have been shown to severely impact the real economy (as discussed in Section \ref{sec:gdp}) and subsequently unemployment \citep{hamilton1983,baumeister2013}. 

Although KRR draws parallels with the first oil crisis in the 1970s and the 2001 recession, its final prediction is nowhere near the realized value. This deficiency can be traced back to several counteracting forces. First, dissimilarities from the 1969-1970 recession, a period marked by rising unemployment rates, contribute to a downward adjustment in the forecast and lead to elevated levels in short positions. Moreover, after rising from low levels during the 1970s and early 1980s, the forecast is brought down again by similarities found with the Great Moderation. 





\vskip 0.15cm
{\noindent \sc \textbf{The Peak for Multiple Steps Ahead (\textcolor{darkgreen2}{$h=2$}, \textcolor{darkred2}{$h=4$})}}. Expectedly, predicting the exceptionally high unemployment rate in 2009Q1 becomes more challenging as $h$ increases. All models deliver substantially lower predictions for $h=\{2,4\}$ compared to $h=1$. While NN and FAAR show strong forecasting performances for both horizons, the tree-based models fail to see similarities with past episodes and produce forecasts near zero.   Given the weaker signals, all models exhibit lower forecast concentration and turnover rates, along with short positions closer to 0 than for $h=1$. 


For our best performing models, NN and FAAR, we observe similar patterns that closely resemble those seen in the one step ahead forecasting exercise, though with reduced intensity.  KRR, on the other hand, predicts only a modest rise in the unemployment rate across all horizons. While the prediction for $h=4$ receives higher contributions from the 1990s and 2001 recessions, these are insufficient to account for the significant impact of the GFC on the labor market.



The tree-based models return forecasts settling approximately at the unconditional mean. LGB struggles to identify any parallels to past periods, particularly for $h=4$. RF, on the other hand, finds some similarities with historical episodes like the early 1990s recession, and to a lesser extent, that of 2001.  However, without detecting proximity to pre-Great Moderation recessions, especially oil crisis  linked with severe upticks in unemployment, these prove insufficient to predict the upcoming peak in the unemployment rate.



\subsection{GDP Growth and Recession Risks in the Post-Covid Era}\label{sec:usrec}

In light of recent policy debates, our final application tackles the challenge of forecasting GDP growth and assessing recession probabilities in the aftermath of the pandemic. We do so by employing two distinct approaches: 1) we rely on our regression techniques discussed in Sections \ref{sec:rr} to \ref{sec:bt} for GDP growth forecasts, and 2) we analyze recession probabilities through extensions to classification models as described in Section \ref{sec:class}. 

\begin{figure}
\caption{\normalsize{Dual Interpretation of Post-Pandemic GDP Growth}} \label{fig:regrec2024}
\centering
\vspace*{-0.5em}
      \includegraphics[width=\textwidth, trim = 0mm 0mm 3mm 0mm, clip]{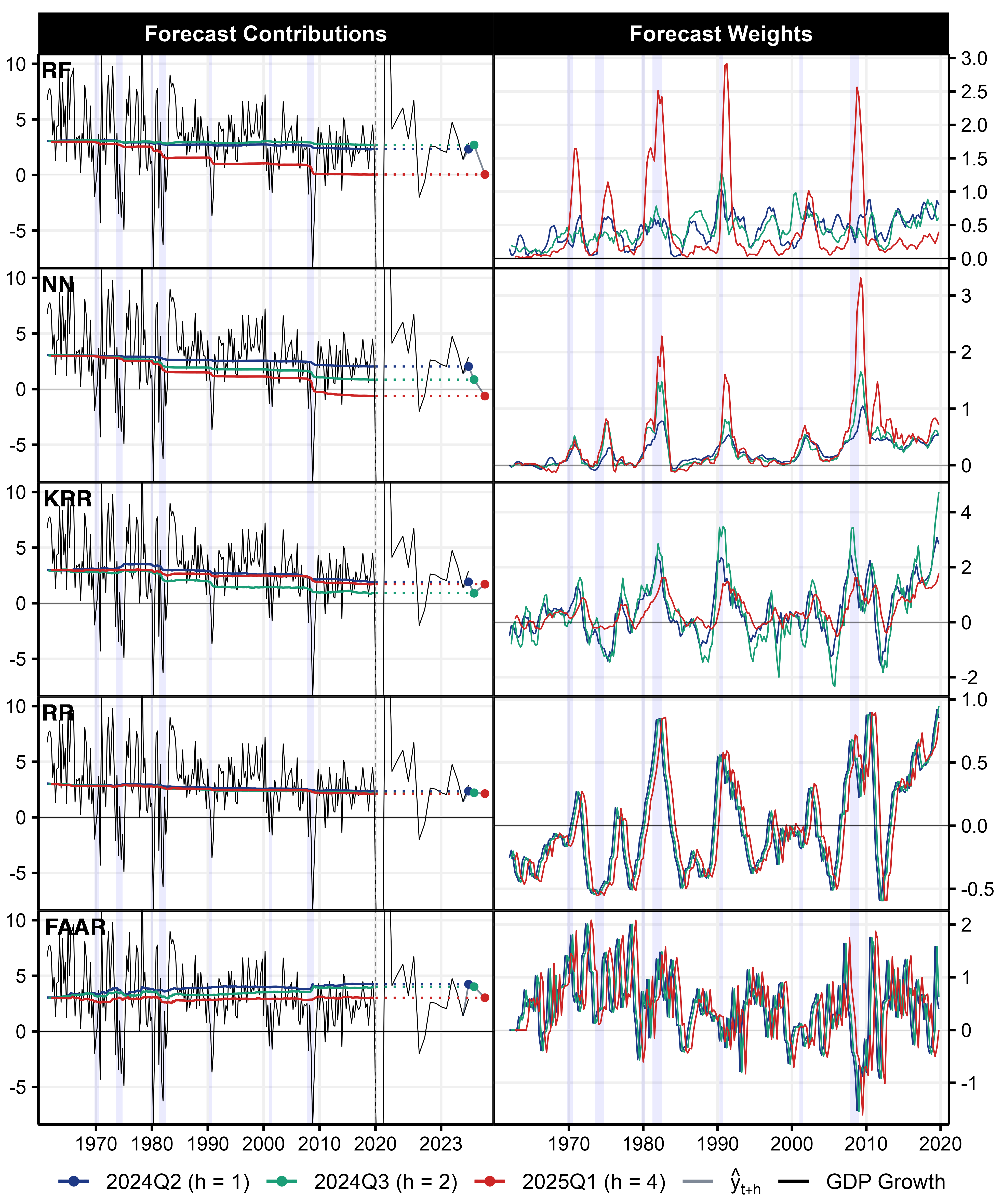}

 \begin{threeparttable}
    \centering
    \vspace*{-1.5em}
    \begin{minipage}{\textwidth}
      \begin{tablenotes}[para,flushleft]
    \setlength{\lineskip}{0.2ex}
    \footscript 
  {\textit{Notes}: The figure presents results from predicting GDP growth one, two, and four steps ahead. The \textbf{left panels} present the \textit{cumulative} sum of forecast contributions $c_{ji}$ over the training sample (1961Q2 to 2019Q4), which collectively sum to the final predicted value $\hat{y}_j$ shown as dots. We initialize $c_{j0}$ at the unconditional average of the sample and present $c_{ji}$ as deviations from this average. The holdout sample ranges from 2020Q1 to 2025Q1, indicated by the dashed line. The \textbf{right panels} show forecast weights $w_j$ as a moving average of four quarters. Lavender shading corresponds to NBER recessions.}
    \end{tablenotes}
  \end{minipage}
  \end{threeparttable}
\end{figure}

For predicting GDP growth, we choose a hold-out sample from 2020Q1 to 2025Q1, and decompose predictions for the end of 2022 and 2024. 
The choice of 2022 is based on growing discussions about recession risks during that year, fueled by intensified monetary tightening in response to exceptionally high inflation. Given concerns about a potential slowdown resurfacing recently, we aim to give an outlook for the coming quarters and provide the models' insights on parallels with past economic conditions in our sample.  Specifically, we provide $h=1$, $h=2$, and, $h=4$ quarters ahead outlooks for a forecaster standing in 2022Q3 and 2024Q1  (see Figures \ref{fig:regrec2022} and \ref{fig:regrec2024}). Therefore, the forecast dates are 2022Q4 and 2024Q2 for $h=1$,  2023Q1 and 2024Q3 for $h=2$, and 2023Q3 and 2025Q1 for $h=4$.

For predicting recession probabilities, we follow the literature and use monthly data, the typical frequency when using the term spread as predictor. Our hold-out sample runs from 2020M1 to 2025M5 and we forecast horizons $h=3$ and $h=12$ starting from three points in time: 2022M5, 2023M5, 2024M5 (see Figure \ref{fig:classrec3} in the appendix and Figure \ref{fig:classrec12}). The set of models for this exercise includes a yield curve (YC) probit model and a nonlinear extension using KRR, both of which take the term spread between the ten-year and three-month Treasury rates as the main predictor \citep{Harvey1989,rudebusch2009}. Moreover, we estimate RF and NN using the variables provided by FRED-MD \citep{mccrackenng} alongside our proxy for the shape of the YC. The target variable is the indicator of US recessions as dated by the NBER.


\vskip 0.15cm
{\noindent \sc \textbf{Predicting GDP Growth}}. Recent GDP growth predictions point to low, or at most, average growth for the end of 2024 and early 2025 (see Figure \ref{fig:regrec2024}). Broadly, our models fall into two camps: the nonlinear models, which identify potential parallels with past recessions and forecast a slowdown in economic activity, and the linear models, which detect fewer signals from turbulent periods in the past, resulting in projections close to the unconditional mean. 

RF and NN produce lowest growth projections for 2025Q1 ($h=4$), which is surprising, as receiving the strongest signals at longer horizons is rather unusual. Both models base their outcomes on similarities with the GFC, the 1990s, and 1980s recessions. Parallels to the GFC can be drawn through financial distress and volatile stock markets, both of which have been shown to be important for forecasting real activity, particularly in relation to downside risks \citep{adrian2019,amburgey2023}. The early 1980s bear similarities to the current situation, as both periods feature rapid and sustained tightening cycles to combat high inflation. Although the US economy has proven robust in recent times, comparable characteristics in terms of speed and duration of the tightening cycle \citep{kwan2023} offer some ground for leveraging similarities. 

On the contrary, the linear models in our set do not indicate contractionary tendencies, but project future growth near the unconditional mean. These predictions can be explained by low weights throughout the training sample, particularly in the case of RR, and offsetting effects within their weighting schemes, especially for FAAR. KRR predicts growth rates somewhat below the unconditional mean for $h=4$, pointing towards a slowdown in real activity in the near term ($h=2$). While KRR identifies some parallels with past recessions, the contributions do not lead to growth falling to zero or turning negative. 


Comparing these findings to predictions in 2022 (see Figure \ref{fig:regrec2022} in the appendix) reveals that, two years earlier, most models predicted zero or negative growth for short horizons, while they pointed to average growth rates for the longer term ($h=4$). This is also reflected in the Survey of Professional Forecasters (SPF), in which projections were considerably revised downwards in the 2022Q3 survey round, projecting a slowdown for 2022Q4 and 2023Q1 followed by a rebound in 2023Q3.\footnote{In the SPF survey round 2022Q3, GDP growth is projected to stand at 1.2\% in 2022Q4, 1.1\% in 2023Q1, and 1.5\% in 2023Q3. Data is retrieved from the FRB Philadelphia's website \href{https://www.philadelphiafed.org/surveys-and-data/real-time-data-research/survey-of-professional-forecasters}{here}.}  Unfortunately, the contraction forecast for 2023Q1, although consensual among our model set, proves to be quite off the mark.  Figure \ref{fig:regrec2022} makes it clear that RF, NN, and to a lesser extent, KRR, overreacted due to their strong emphasis on the GFC and past rapid monetary tightening cycles. In contrast, RR and FAAR inferences were more subtle, drawing on slower-growth periods like the post-GFC years, resulting in near-zero growth forecasts.

{\color{black} A technical clarification is in order for the linear models shown in Figures \ref{fig:regrec2022} and \ref{fig:regrec2024}, where their weights are visibly translations of one another across all horizons. For linear models such as FAAR and RR (assuming similar hyperparameters for RR), the proximity weights time series have to be identical across horizons, differing only by a shift of $h$ steps ahead. This is due to the identical feature space used for all targets. Mathematically, $ \boldsymbol{w}_j \equiv X_j \boldsymbol{X}’ (\boldsymbol{X}\boldsymbol{X}’ + \lambda {I}_N)^{-1} $ remains unchanged as long as $X_j$ and $\boldsymbol{X}$ are unchanged, even when $\boldsymbol{y}$ varies. In contrast, nonlinear models optimize features uniquely for each target, resulting in different proximity interpretations for the same $X_j$ and $\boldsymbol{X}$ pair.}

\vskip 0.15cm
{\noindent \sc \textbf{Predicting Recession Probabilities}}. Discussions about the US economy facing a recession have been ongoing since the start of the tightening cycle in 2022, though they intensified recently, particularly after the disappointing August 2024 jobs report, which triggered the Sahm rule \citep{sahm2019} to signal heightened recession risks. By targeting recessions directly, our approach facilitates a nuanced understanding of the predicted probabilities of these tail events occuring, which is based on contributions from past historical episodes in our sample. Figure \ref{fig:classrec12} shows that, overall, our models agree on low to zero recession probabilities for mid-2023 (as predicted in 2022 for 12 steps ahead). For mid-2024 and mid-2025, on the other hand, the models indicate elevated risks, drawing clear parallels to past periods of economic contraction.

YC and KRR both indicate high recession probabilities for 2024M5 and lower yet clearly elevated risks for 2025M5. While YC's predictions feature constant upward bound contributions, KRR identifies more distinct drivers, suggesting a cautious, potentially alarmist outlook. For 2024M5, recession probabilities are pushed above 90\% due to the impact of the recession following the first oil crisis, which is partially offset by moderating influence of the early 2000s. For 2025M5, on the other hand, largest contributions come from the 1990s recession and the GFC. 

RF and NN predict highest recession probabilities when using most recent data and forecasting 2025M5. RF's predictions are shaped by stepwise contributions from each major recession in the sample, with the early 1980s and the GFC having the most significant impact. This aligns with the findings from recent GDP growth predictions and suggests that prevailing conditions resemble those of past economic disruptions, particularly the strong tightening cycle of the 1980s and tight financial conditions observed during both the early 1980s and the GFC. The signals received lead to over 50\% recession probability for 2024M5 and almost 75\% for 2025M5. NN's predictions are driven by moderate contributions over the training sample, resulting in low recession probabilities for mid-2023 and mid-2024 (0\% and 10\%, respectively), and 33\% in 2025M5. 


\begin{figure}[t!]
\caption{\normalsize{Dual Interpretation of Post-Pandemic Recession Probabilities ($h=12$)}} \label{fig:classrec12}
\centering
\vspace*{-0.5em}
      \includegraphics[width=\textwidth, trim = 0mm 0mm 3mm 0mm, clip]{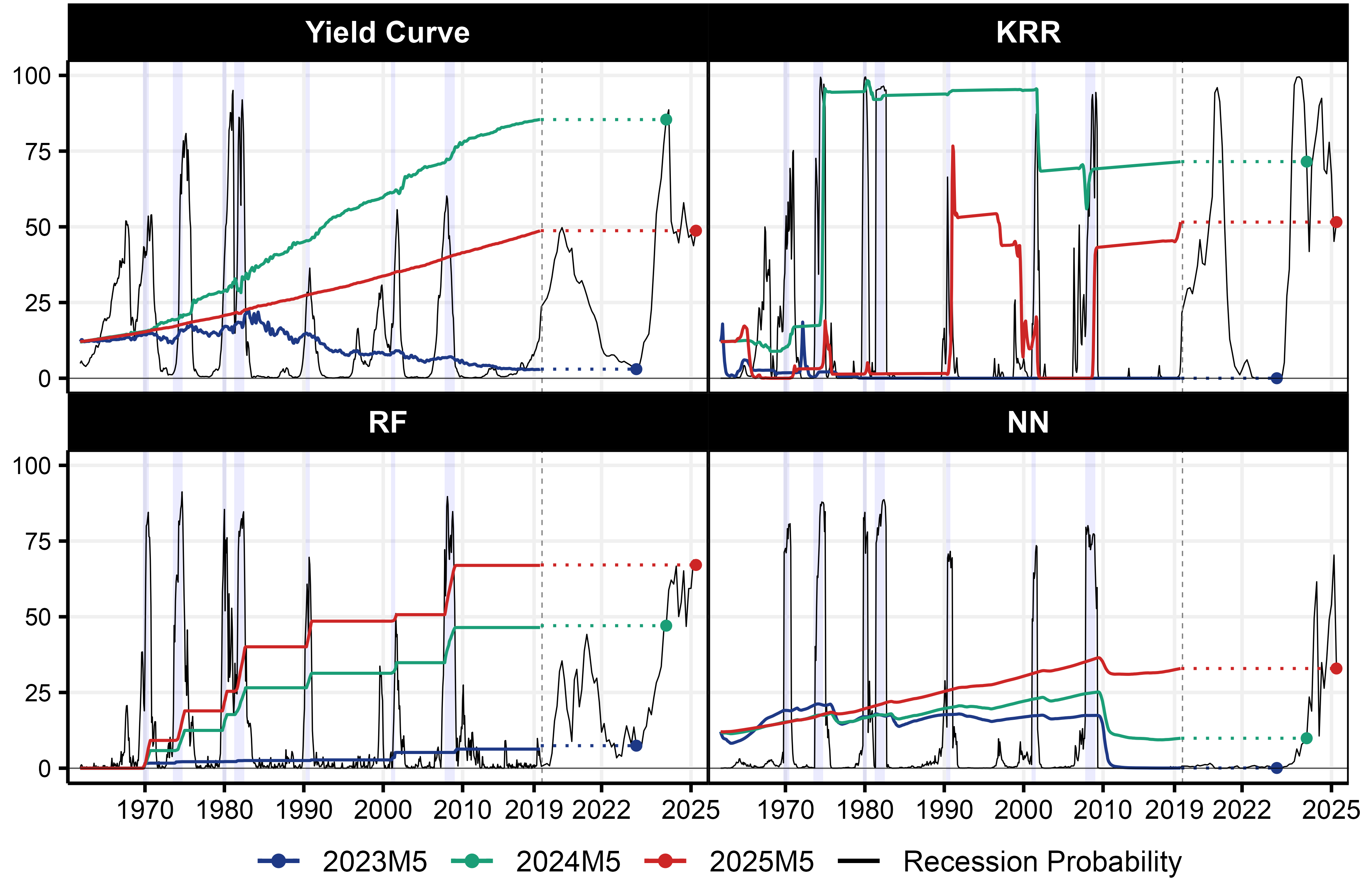}

 \begin{threeparttable}
    \centering
    \vspace*{-1.5em}
    \begin{minipage}{\textwidth}
      \begin{tablenotes}[para,flushleft]
    \setlength{\lineskip}{0.2ex}
    \footscript 
  {\textit{Notes}: The figure presents results from predicting recession probabilities 12 steps ahead. We present the \textit{cumulative} sum of forecast contributions $c_{ji}$ over the training sample (1961M4 to 2019M12), which collectively sum to the final predicted value $\hat{y}_j$ shown as dots. We initialize $c_{j0}$ at the unconditional average of the sample and present $c_{ji}$ as deviations from this average. Up to 2020M1 we plot in-sample results and show out-of-sample predictions through to 2025M5, indicated by the dashed line. Lavender shading corresponds to NBER recessions, i.e., the target variable.}
    \end{tablenotes}
  \end{minipage}
  \end{threeparttable}
\end{figure}

On a more technical note, one might wonder why YC’s contributions resemble straight lines, while RF appears more like a staircase. This pattern arises from the dichotomous nature of the forecasting target and the presence or absence of a short position. By design, expansions in RF may receive some weight, but their contribution remains zero, as the target is encoded as $y_i \in \{0,1\}$. If $y_i$ were encoded as $\{-1,1\}$ instead, these periods would look like straight downward lines. In 0-1 encoding, the impact of “expansion contributions” shrinks the relative share of recession contributions, resulting in a mostly flat line with few (and muted) positive contributions, such as that seen for 2023M5.

 A similar reasoning explains YC’s upward slopes in 2024M4 and 2025M5, where contributions come from both positive recession contributions and the “negatives” of expansions. The latter is impossible in RF, which features no short position. In the context of the YC model, the input data used for training consists of only a single indicator. Because of this limited input space, the model has little information to distinguish between different recessions within the training sample. Essentially, all recessions appear similar to the model as they can only be differentiated by the intensity of yield curve inversions. As a result, they all contribute more or less equally to the out-of-sample prediction, or they do not.

For the short horizon ($h=3$, presented in Figure \ref{fig:classrec3} in the appendix), models agree on low to modest probabilities of a recession occurring, with the most recent data carrying somewhat elevated risks. These risks are mainly driven by the early 1990s recession (for KRR) and the 2000s recession (for RF and KRR). Notably, KRR distinguishes itself from the other models by predicting alarmingly high recession risks for 2023M8, reaching 89\%. Interestingly, and reinforcing our previous findings, it identifies clear parallels with the twin recessions.


Overall, a comparison of Figures \ref{fig:regrec2024} and \ref{fig:classrec12} shows that KRR, RF, and NN point in similar qualitative directions with their predictions of recession probabilities and GDP growth. For 2024Q2, higher growth projections are supported by lower recession probabilities, while for 2025Q1, lower growth projections align with increased recession odds in 2025M5.
 
 





\section{Conclusion  }\label{sec:con} 


We introduced a novel approach to interpreting machine learning forecasts by decomposing out-of-sample predictions into contributions from individual training observations. The associated data portfolio weights can be seen as proximity scores between current economic conditions and those in the estimation sample. This dual decomposition method proves especially valuable in macroeconomic forecasting, where datasets often feature numerous predictors but limited observations. By visualizing these contributions as time series and analyzing them through portfolio measures such as forecast concentration, we offer a new angle from which to open the black box. Given the widespread use of “judgement” in economic forecasting, often based on historical analogies, the tools developed in this paper offer a bridge for the integration of narrative insights into dense quantitative methods.


There are many avenues for future research. One promising direction is to adopt a more proactive approach with data portfolio weights, moving beyond observation and analysis. For example, sparsity tools could be applied to refine these weights further, as discussed for random forest-based probabilities in \cite{koster2024simplifying}. More broadly, regularization techniques (or priors in Bayesian frameworks) could be developed specifically for proximity coefficients rather than the usual regression parameters. This could not only ease interpretation, but also improve forecasts themselves.
 







\clearpage

\setlength\bibsep{5pt}
               
\bibliographystyle{apalike}
 
\setstretch{0.75}

 \def\dboxpath{/UQAM} 

\bibliography{ref_pgc_v181204}

\clearpage
 
\appendix
\newcounter{saveeqn}
\setcounter{saveeqn}{\value{section}}
\renewcommand{\theequation}{\mbox{\Alph{saveeqn}.\arabic{equation}}} \setcounter{saveeqn}{1}
\setcounter{equation}{0}
\setstretch{1.25}
 
\pagebreak
 
 
\appendix
 
\section{Appendix}

\subsection{Hemisphere Neural Network}\label{sec:HNNprimer}

For forecasting inflation, we incorporate the Hemisphere Neural Network (HNN) from \cite{HNN} into our set of models. It features a specific structure based on a nonlinear Phillips curve. We define seven hemispheres ($\eta_m$), where $\eta_{LR}(t)$ stands for an estimate of long-run expectations (LR), estimated only with information from a linear time-trend $t$. The potentially time-varying effect of the output gap (G) instead is a combination of (i) an estimate of G itself ($\eta_{G} (\mathcal{H}_{G} \setminus t)$), which is the output of an NN fed with information $\mathcal{H}_{G}$ not containing the time-trend $t$, and (ii) the loading ($\eta_{\zeta_{G}}(t)$) which is again the output of an NN, but only fed with information about $t$. Commodity prices (C) and short-run expectations (SR) enter the final layer in the same manner as G. This version of the HNN refers to \cite{HNN}'s factorized version of the model (HNN-F). In terms of a more formal notation, this results in

\begin{equation}
\hat{y}_j = \eta_{LR}(t) + \eta_{\zeta_{SR}}(t) \eta_{SR}(\mathcal{H}_{SR} \setminus t) + \eta_{\zeta_{G}}(t) \eta_{G} (\mathcal{H}_{G} \setminus t) + \eta_{\zeta_{C}}(t) \eta_{C}(\mathcal{H}_{C} \setminus t).
\end{equation}

Each hemisphere is estimated as a standard feed-forward neural network as described in Section \ref{sec:nn}. Note that in this case we define $Z^{\eta_m}_{L-1,j}  \equiv \Psi(\eta_{m_j})$ where $\eta_{m_j}$ denotes the $j^{th}$ observation of hemisphere $\eta_m$.

The network structure includes three layers with 400 neurons per hemisphere. Time-variation is captured by adding trends, which are modeled with three layers and 100 neurons each. We set the maximum number of \texttt{epochs} to 500, the \texttt{learning.rate} to 0.05, and the \texttt{dropout.rate} to 0.2. For early stopping we use a subset of 65\% of the training data to estimate the parameters and use the remaining 35\% to determine when to stop the optimization process. The tolerance parameter is 0.01 and the size of the blocks to shuffle is eight quarters. The number of bootstrap is 50. The loss function is based on the mean squared error and the optimizer is Adam.

Finally, we list the variables (in FRED-QD mnemonics) that enter each hemisphere:

\begin{lstlisting}[language=R]
real.activity.hemisphere <- c("PAYEMS","USPRIV","MANEMP","SRVPRD",
                  "USGOOD" ,"DMANEMP","NDMANEMP","USCONS","USEHS",
                  "USFIRE","USINFO","USPBS","USLAH","USSERV",
                  "USMINE","USTPU","USGOVT","USTRADE",
                  "USWTRADE","CES9091000001","CES9092000001",
                  "CES9093000001","CE16OV","CIVPART",
                  "UNRATE","UNRATESTx","UNRATELTx","LNS14000012",
                  "LNS14000025","LNS14000026",
                  "UEMPLT5","UEMP5TO14","UEMP15T26","UEMP27OV",
                  "LNS13023621","LNS13023557",
                  "LNS13023705","LNS13023569","LNS12032194",
                  "HOABS","HOAMS","HOANBS","AWHMAN",
                  "AWHNONAG","AWOTMAN","HWIx","UEMPMEAN",
                  "CES0600000007", "HWIURATIOx","CLAIMSx","GDPC1",
                  "PCECC96","GPDIC1","OUTNFB","OUTBS","OUTMS",
                  "INDPRO","IPFINAL","IPCONGD","IPMAT","IPDMAT",
                  "IPNMAT","IPDCONGD","IPB51110SQ","IPNCONGD",
                  "IPBUSEQ","IPB51220SQ","TCU","CUMFNS",
                  "IPMANSICS","IPB51222S","IPFUELS") 
                  
SR.expec.hemisphere <- c("Y", "PCECTPI","PCEPILFE",
                  "GDPCTPI","GPDICTPI","IPDBS", "CPILFESL","CPIAPPSL",
                  "CPITRNSL","CPIMEDSL","CUSR0000SAC","CUSR0000SAD",
                  "WPSFD49207",    "PPIACO","WPSFD49502","WPSFD4111",
                  "PPIIDC","WPSID61","WPSID62","CUSR0000SAS","CPIULFSL",
                  "CUSR0000SA0L2","CUSR0000SA0L5", "CUSR0000SEHC",
                  "spf_cpih1","spf_cpi_currentYrs","inf_mich")
                  
commodities.hemisphere <- c("WPU0531","WPU0561","OILPRICEx","PPICMM")

LR.expec.hemisphere <- c("trend")


\end{lstlisting}

\subsection{Additional Graphs and Tables}\label{sec:addresults}

\begin{figure}[H]
\caption{\normalsize{Moving Average Forecast Contributions for Inflation ($h=1$, all models)}} \label{fig:contribma4_infl}
\centering
\vspace*{-0.5em}

 \includegraphics[width=\textwidth, trim = 0mm 0mm 0mm 0mm, clip]{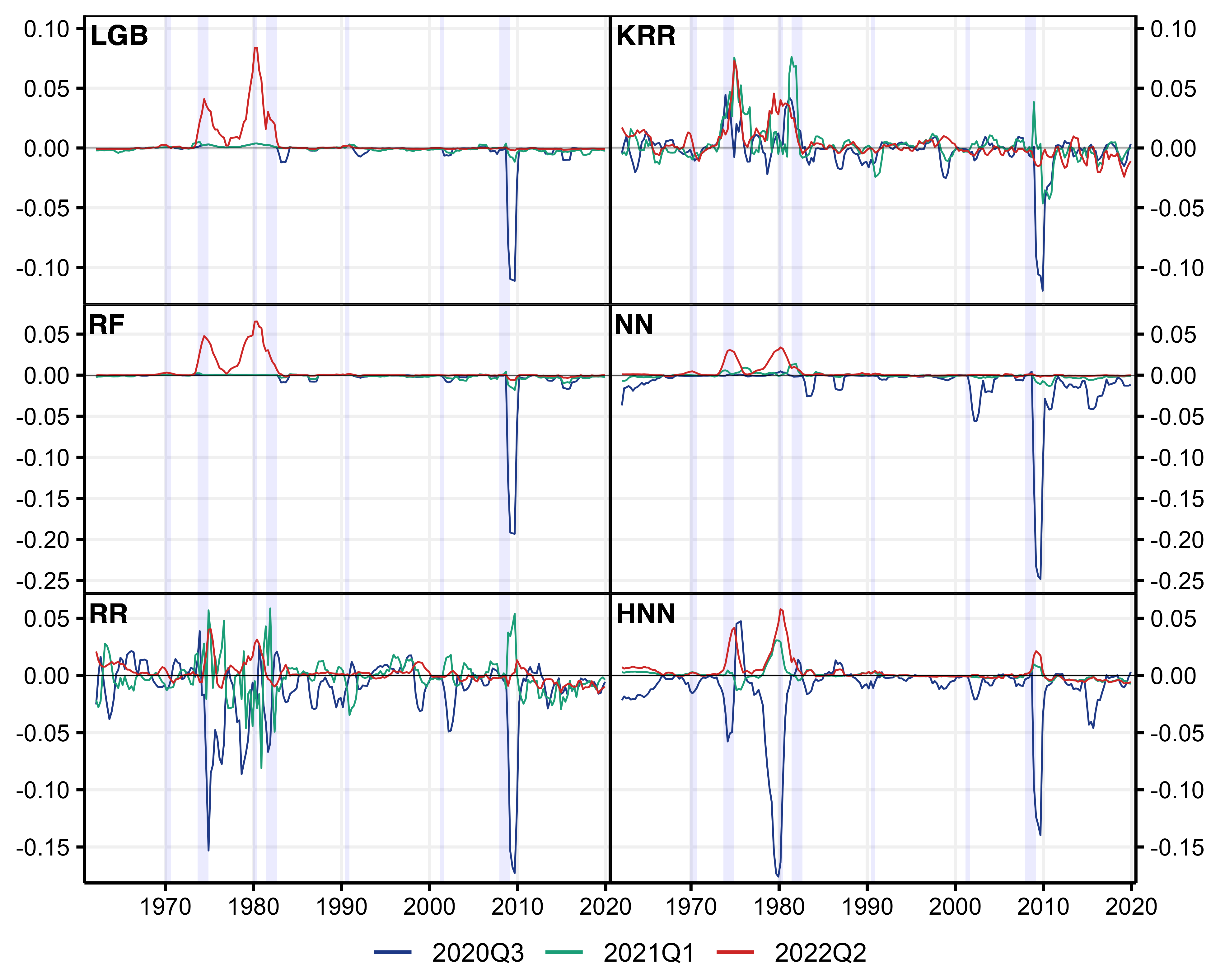}

 \begin{threeparttable}
    \centering
    \begin{minipage}{\textwidth}
      \begin{tablenotes}[para,flushleft]
    \setlength{\lineskip}{0.2ex}
    \footscript 
  {\textit{Notes}: The figure presents $c_{ji}$ as a moving average of four quarters. Lavender shading corresponds to NBER recessions.}
    \end{tablenotes}
  \end{minipage}
  \end{threeparttable}
\end{figure}

\begin{figure}[H]
\caption{\normalsize{Moving Average Forecast Contributions for GDP Growth ($h=1$)}} \label{fig:contribma4_gdp}
\centering 
\vspace*{-0.5em}

 \includegraphics[width=\textwidth, trim = 0mm 0mm 0mm 0mm, clip]{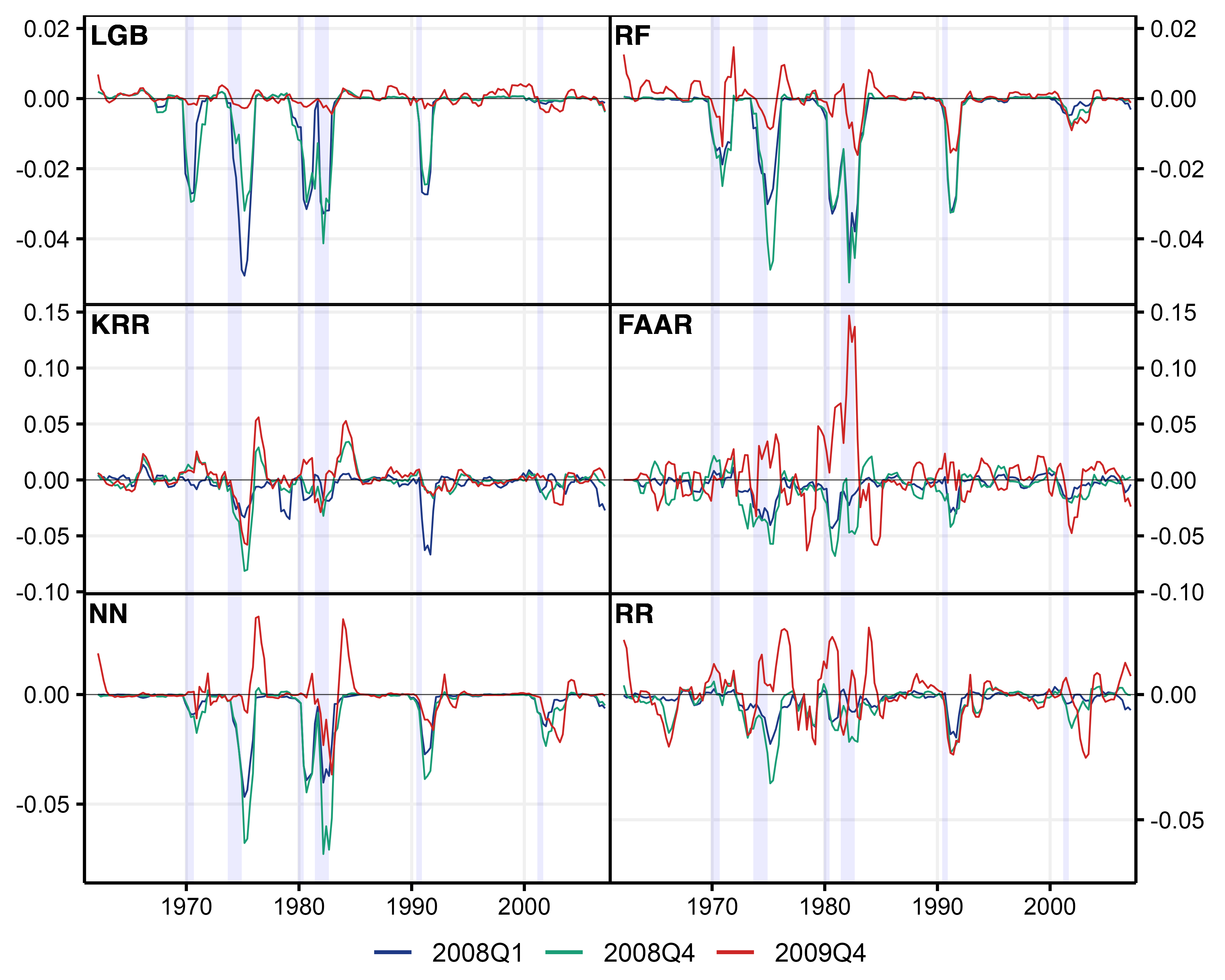}

 \begin{threeparttable}
    \centering
    \begin{minipage}{\textwidth}
      \begin{tablenotes}[para,flushleft]
    \setlength{\lineskip}{0.2ex}
    \footscript 
  {\textit{Notes}: The figure presents $c_{ji}$ as a moving average of four quarters. Lavender shading corresponds to NBER recessions.}
    \end{tablenotes}
  \end{minipage}
  \end{threeparttable}
\end{figure}

\begin{figure}[H]
\caption{\normalsize{Moving Average Forecast Contributions for Unemployment ($h=1$)}} \label{fig:contribma4_unemp}
\centering
\vspace*{-0.5em}

 \includegraphics[width=\textwidth, trim = 0mm 0mm 0mm 0mm, clip]{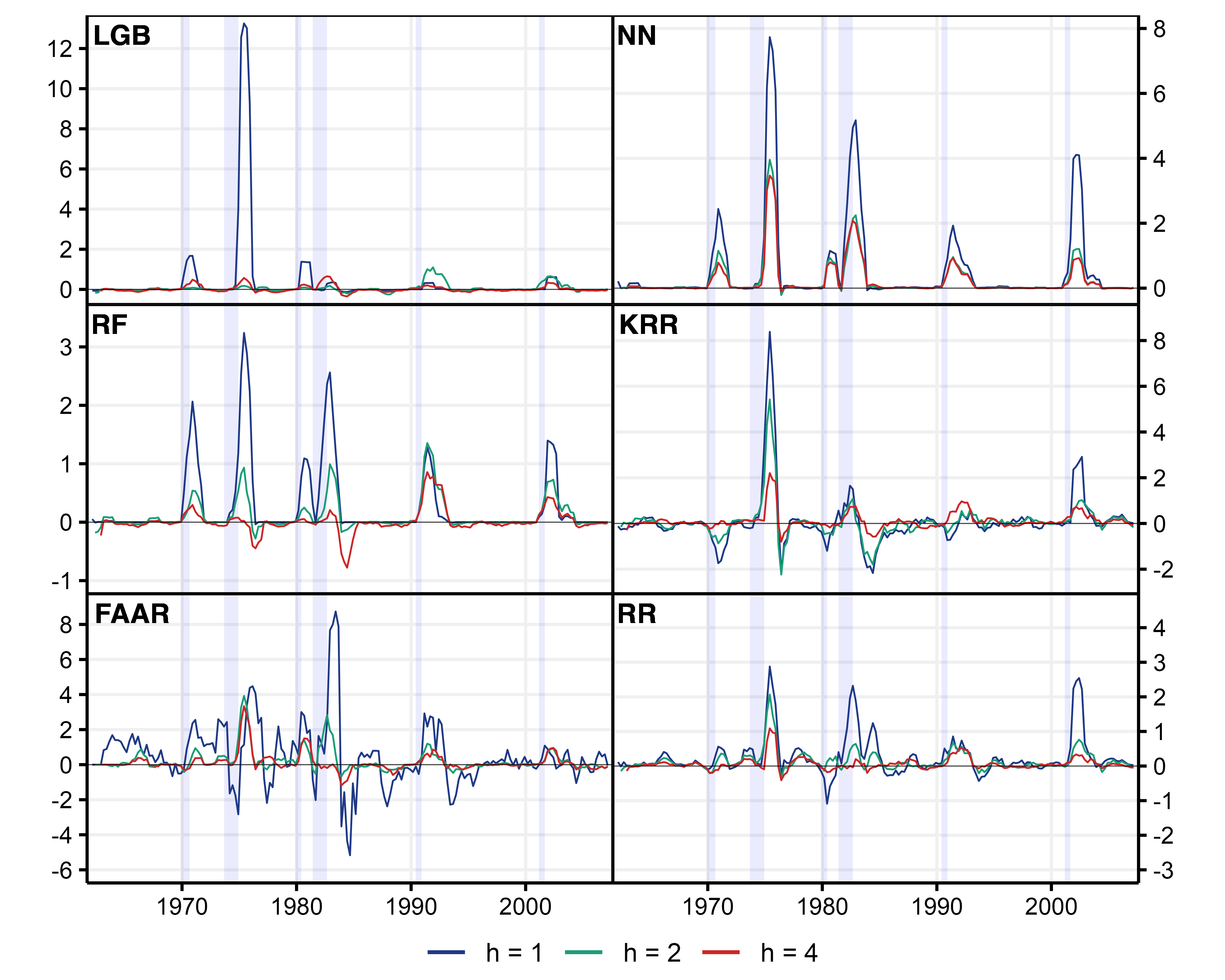}

 \begin{threeparttable}
    \centering
    \begin{minipage}{\textwidth}
      \begin{tablenotes}[para,flushleft]
    \setlength{\lineskip}{0.2ex}
    \footscript 
  {\textit{Notes}: The figure presents $c_{ji}$ as a moving average of four quarters. Lavender shading corresponds to NBER recessions.}
    \end{tablenotes}
  \end{minipage}
  \end{threeparttable}
\end{figure}

\begin{table}[H]
\vspace{1em}
  \footnotesize
  \centering
  \begin{threeparttable}
  \caption{\normalsize {Forecast Statistics for GDP Growth ($h=1$)} \label{tab:metrics_gdp}
    \vspace{-0.3cm}}
    \setlength{\tabcolsep}{0.6em} 
    \setlength\extrarowheight{2.9pt}

    \begin{tabular}{l| rrr|rrr|rrr|r}
    \toprule \toprule
    \addlinespace[2pt]
    \multicolumn{1}{l|}{} & \multicolumn{3}{c|}{Concentration}  & \multicolumn{3}{c|}{Leverage} & \multicolumn{3}{c|}{Short Position} & \multicolumn{1}{c}{Turnover}  \\ 
    \cmidrule(lr){2-4} \cmidrule(lr){5-7} \cmidrule(lr){8-10} \cmidrule(lr){11-11}
    \multicolumn{1}{r|}{} & \multicolumn{1}{c}{2008Q1}  & \multicolumn{1}{c}{2008Q4} & \multicolumn{1}{c|}{2009Q4} & \multicolumn{1}{c}{2008Q1}  & \multicolumn{1}{c}{2008Q4} & \multicolumn{1}{c|}{2009Q4}  & \multicolumn{1}{c}{2008Q1}  & \multicolumn{1}{c}{2008Q4} & \multicolumn{1}{c|}{2009Q4} & \multicolumn{1}{c}{Overall} \\ 
    \midrule
  LGB & 0.54 & 0.53 & 0.20 & 1.00 & 1.00 & 1.00 & 0 & 0 & 0 & 8.70 \\ 
  RF & 0.32 & 0.35 & 0.23 & 1.00 & 1.00 & 1.00 & 0 & 0 & 0 & 5.04 \\ 
  NN & 0.37 & 0.32 & 0.25 & 0.86 & 1.34 & 1.73 & -0.03 & -0.04 & -0.06 & 6.39 \\ 
  KRR & 0.19 & 0.19 & 0.21 & 1.00 & 1.00 & 1.00 & -0.99 & -1.29 & -1.54 & 38.09 \\ 
  RR & 0.16 & 0.17 & 0.17 & 1.00 & 1.00 & 1.00 & -0.53 & -1.01 & -1.66 & 21.91 \\
  FAAR & 0.16 & 0.16 & 0.16 & 1.00 & 1.00 & 1.00 & -0.84 & -1.62 & -3.79 & 97.74 \\ 
   \bottomrule \bottomrule 
\end{tabular}
\begin{tablenotes}[para,flushleft]
  \footscript 
    \textit{Notes}: The table summarizes forecast metrics as discussed in Section \ref{sec:derivatives}. Forecast concentration shows the proportion of total weights attributed to the top 5\% of the weights ($Q=5$).
  \end{tablenotes}
\end{threeparttable}
\end{table}

\begin{table}[H]
\vspace{1em}
  \footnotesize
  \centering
  \begin{threeparttable}
  \caption{\normalsize {Forecast Statistics for Unemployment (2009Q1)} \label{tab:metrics_unemp}
    \vspace{-0.3cm}}
    \setlength{\tabcolsep}{0.9em} 
    \setlength\extrarowheight{2.9pt}

    \begin{tabular}{l| rrr|rrr|rrr|rrr}
    \toprule \toprule
    \addlinespace[2pt]
    \multicolumn{1}{l|}{} & \multicolumn{3}{c|}{Concentration}  & \multicolumn{3}{c|}{Leverage} & \multicolumn{3}{c|}{Short Position} & \multicolumn{3}{c}{Turnover}  \\ 
    \cmidrule(lr){2-4} \cmidrule(lr){5-7} \cmidrule(lr){8-10} \cmidrule(lr){11-13}
    \multicolumn{1}{r|}{$h \rightarrow$} & \multicolumn{1}{c}{$1$}  & \multicolumn{1}{c}{$2$} & \multicolumn{1}{c|}{$4$} & \multicolumn{1}{c}{$1$} & \multicolumn{1}{c}{$2$} & \multicolumn{1}{c|}{$4$} & \multicolumn{1}{c}{$1$} & \multicolumn{1}{c}{$2$} & \multicolumn{1}{c|}{$4$} & \multicolumn{1}{c}{$1$} & \multicolumn{1}{c}{$2$} & \multicolumn{1}{c}{$4$} \\ 
    \midrule
  LGB & 0.63 & 0.34 & 0.10 & 1.00 & 1.00 & 1.00 & 0 & 0 & 0 & 6.00 & 5.91 & 0.52 \\ 
  RF & 0.42 & 0.30 & 0.22 & 1.00 & 1.00 & 1.00 & 0 & 0 & 0 & 5.04 & 4.39 & 5.26 \\ 
  NN & 0.38 & 0.31 & 0.34 & 2.06 & 1.06 & 0.87 & -0.07 & -0.02 & -0.03 & 7.52 & 4.64 & 7.04 \\ 
  KRR & 0.18 & 0.19 & 0.20 & 1.00 & 0.99 & 0.81 & -2.02 & -1.50 & -0.70 & 52.21 & 42.95 & 29.23 \\ 
  RR & 0.17 & 0.17 & 0.17 & 1.00 & 1.00 & 1.00 & -1.70 & -1.01 & -0.65 & 21.59 & 19.68 & 14.60 \\ 
  FAAR & 0.14 & 0.15 & 0.16 & 1.00 & 1.00 & 1.00 & -8.28 & -1.33 & -0.87 & 108.71 & 93.43 & 53.55 \\ 
   \bottomrule \bottomrule 
\end{tabular}
\begin{tablenotes}[para,flushleft]
  \footscript 
    \textit{Notes}: The table summarizes forecast metrics as discussed in Section \ref{sec:derivatives}. Forecast concentration shows the proportion of total weights attributed to the top 5\% of the weights ($Q=5$).
  \end{tablenotes}
\end{threeparttable}
\end{table}

\begin{figure}[H]
\caption{\normalsize{Overall Historical Importance}} \label{fig:ohi}
\centering
\vspace*{-0.5em}

\begin{minipage}{0.49\textwidth}
 \includegraphics[scale=0.4, trim = 4mm 0mm 0mm 0mm, clip]{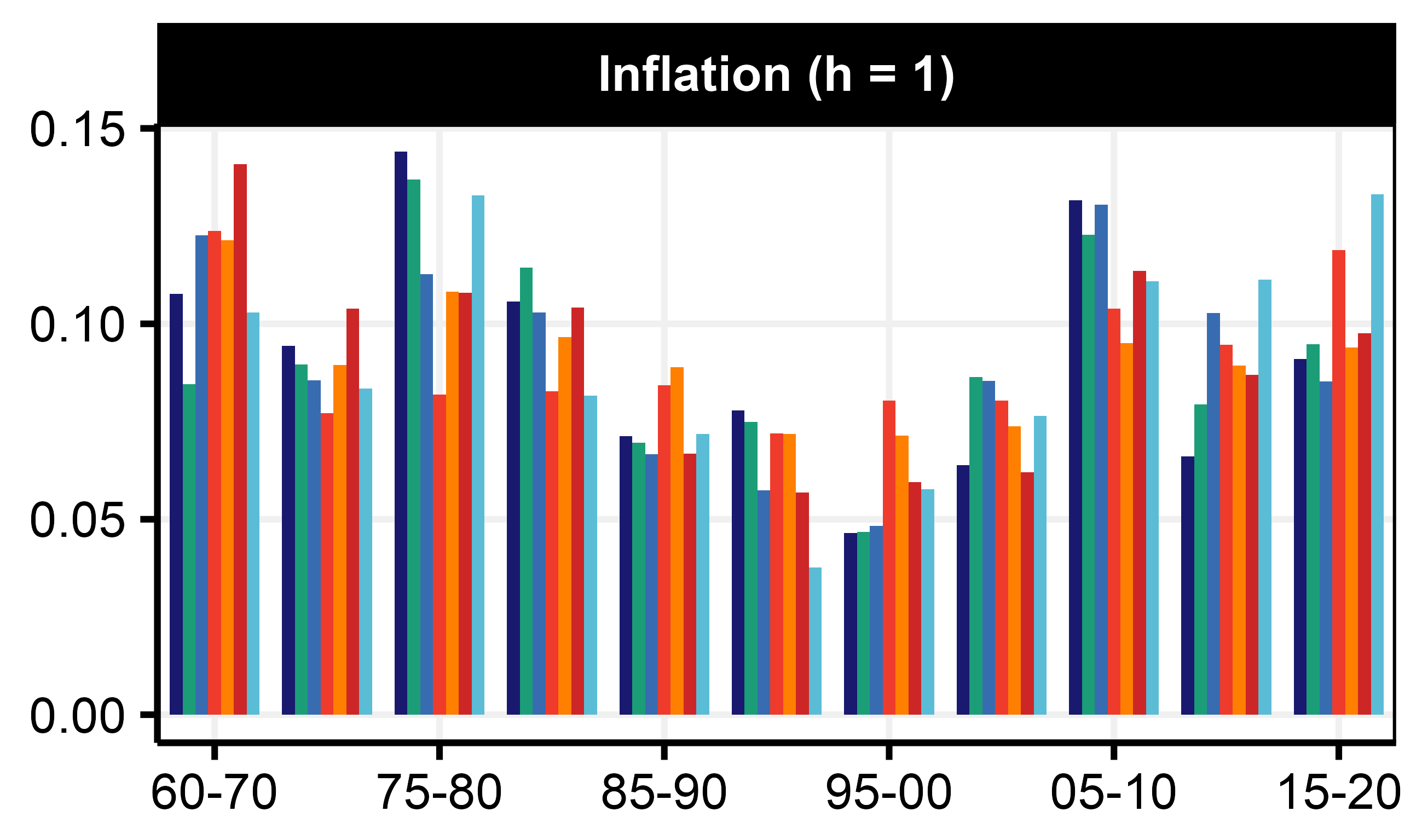}
\end{minipage}
\begin{minipage}{0.49\textwidth}
 \includegraphics[scale=0.4, trim = -2mm 0mm 0mm 0mm, clip]{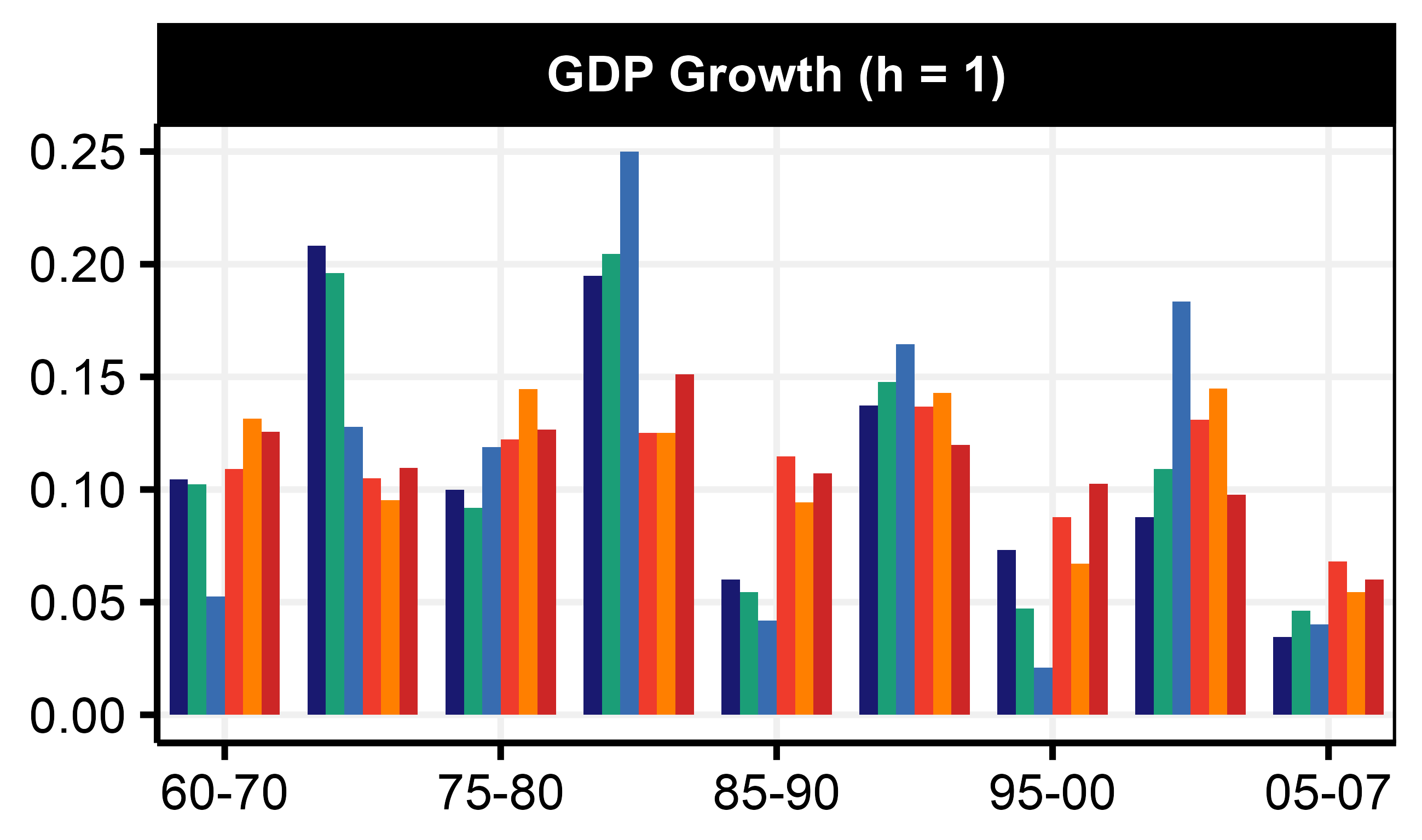}
\end{minipage}

\begin{minipage}{\textwidth}
 \includegraphics[width=\textwidth, trim = -2mm 0mm 0mm 0mm, clip]{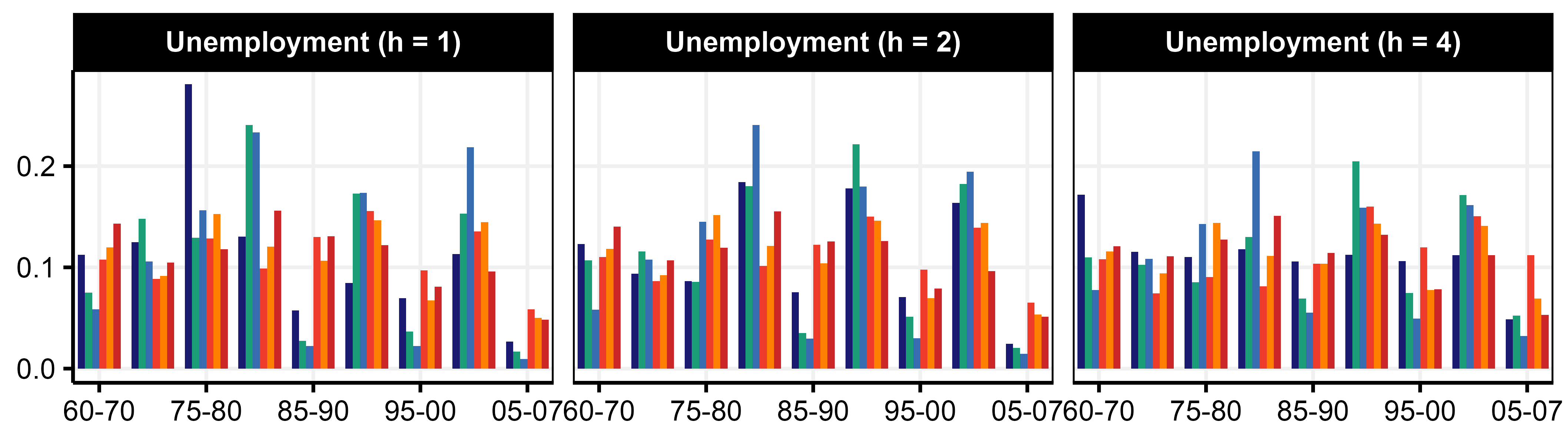}
\end{minipage}

\begin{minipage}{\textwidth}
\centering
 \includegraphics[width=0.7\textwidth, trim = 0mm 0mm 0mm 0mm, clip]{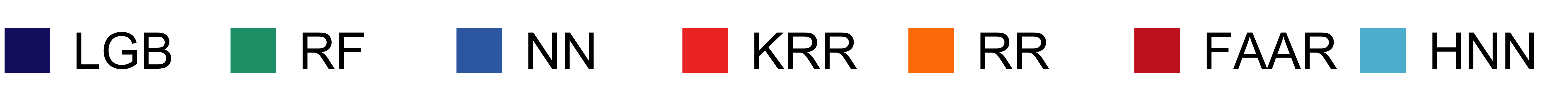}
\end{minipage}

 \begin{threeparttable}
    \centering
    \begin{minipage}{\textwidth}
      \begin{tablenotes}[para,flushleft]
    \setlength{\lineskip}{0.2ex}
    \footscript 
  {\textit{Notes}: The figure presents overall historical importance as discussed in Section \ref{sec:derivatives}. To assess the importance of different historical episodes, we segment our training sample into 5-year intervals. For illustrative purposes, the first bucket bundles all available months of the 1960s.}
    \end{tablenotes}
  \end{minipage}
  \end{threeparttable}
\end{figure}

 \begin{table}[H]
  \footnotesize
  \centering
  \begin{threeparttable}
  \caption{\normalsize {Point Forecasting Performance} \label{tab:rmse}
    \vspace{-0.3cm}}
    \setlength{\tabcolsep}{1.5em} 
    \setlength\extrarowheight{2.9pt}

    \begin{tabular}{l| rrrrrrr}
    \toprule \toprule
    \addlinespace[2pt]
    \multicolumn{1}{l|}{} & \multicolumn{1}{c}{FAAR}  & \multicolumn{1}{c}{KRR} & \multicolumn{1}{c}{LGB} & \multicolumn{1}{c}{NN} & \multicolumn{1}{c}{RF} & \multicolumn{1}{c}{RR} & \multicolumn{1}{c}{HNN}  \\ 
    \midrule
    \rowcolor{gray!15} 
    \multicolumn{8}{l}{Inflation ($h=1$)}   \\ \addlinespace[2pt]
    2020Q1-2024Q1 & 4.30 & 0.90 & 0.80 & 1.50 & 0.98 & 1.60 & 1.36 \\ 
    2021Q1-2024Q1 & 1.82 & 0.97 & 0.99 & 1.35 & 1.04 & 1.45 & 0.90 \\ 
    \addlinespace[5pt] 
    \rowcolor{gray!15} 
    \multicolumn{8}{l}{GDP Growth ($h=1$)}   \\ \addlinespace[2pt]
    2007Q2-2009Q4 & 0.63 & 1.11 & 0.90 & 0.63 & 0.78 & 0.85 & -- \\ 
    2020Q1-2024Q2 & 1.32 & 0.95 & 0.89 & 0.98 & 0.88 & 0.95 & -- \\
    2021Q1-2024Q2 & 0.96 & 0.91 & 0.77 & 0.99 & 0.82 & 0.75 & -- \\ 
    \addlinespace[5pt] 
    \rowcolor{gray!15} 
    \multicolumn{8}{l}{GDP Growth ($h=2$)}   \\ \addlinespace[2pt]
    2020Q1-2024Q2 & 1.16 & 0.94 & 0.94 & 0.95 & 0.94 & 0.94 & -- \\
    2021Q1-2024Q2 & 1.88 & 0.85 & 0.84 & 0.97 & 0.83 & 0.69 & -- \\ 
    \addlinespace[5pt] 
    \rowcolor{gray!15} 
    \multicolumn{8}{l}{GDP Growth ($h=4$)}   \\ \addlinespace[2pt]
    2020Q1-2024Q2 & 0.97 & 0.95 & 0.96 & 0.95 & 0.97 & 0.96 & -- \\
    2021Q1-2024Q2 & 0.77 & 0.55 & 0.54 & 0.55 & 0.59 & 0.52 & -- \\ 
    \addlinespace[5pt] 
    \rowcolor{gray!15} 
    \multicolumn{8}{l}{$\Delta$ Unemployment (2007Q2-2009Q4)}   \\ \addlinespace[2pt]
    $h = 1$ & 0.70 & 1.54 & 0.84 & 0.78 & 0.94 & 1.08 & -- \\ 
    $h = 2$ & 0.90 & 1.16 & 1.10 & 0.69 & 0.98 & 0.97 & -- \\ 
    $h = 4$ & 0.85 & 0.88 & 0.99 & 0.91 & 0.93 & 0.96 & -- \\ 
   \bottomrule \bottomrule 
\end{tabular}
\begin{tablenotes}[para,flushleft]
  \footscript 
    \textit{Notes}: This table shows root mean squared errors (RMSE) against an autoregressive model of order 4.
  \end{tablenotes}
\end{threeparttable}
\end{table}

\begin{figure}[H]
\caption{\normalsize{Dual Interpretation of Post-Pandemic GDP Growth}} \label{fig:regrec2022}
\centering
\vspace*{-0.5em}
      \includegraphics[width=\textwidth, trim = 0mm 0mm 3mm 0mm, clip]{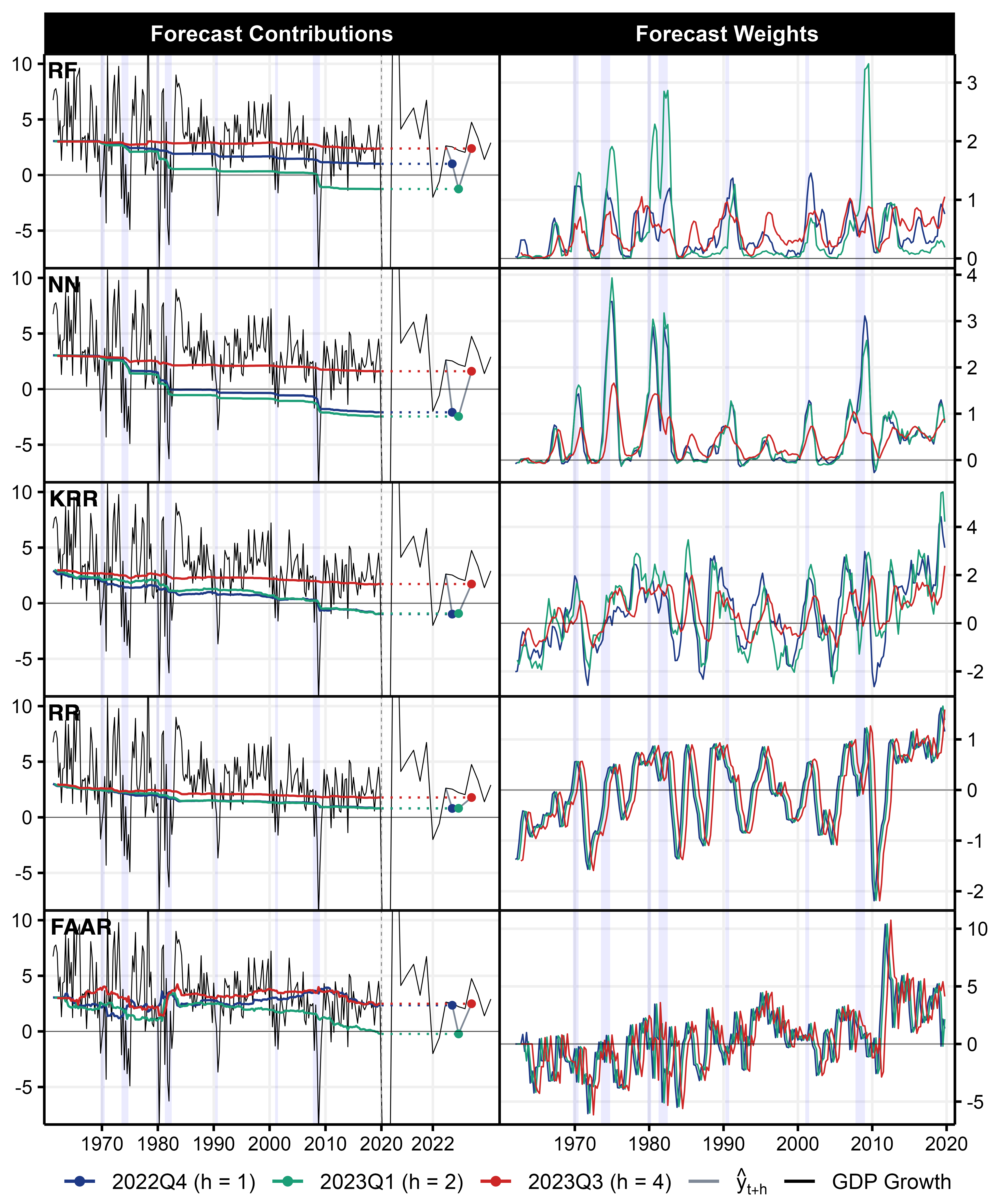}

 \begin{threeparttable}
    \centering
    \vspace*{-1.2em}
    \begin{minipage}{\textwidth}
      \begin{tablenotes}[para,flushleft]
    \setlength{\lineskip}{0.2ex}
    \footscript 
  {\textit{Notes}: The figure presents results from predicting GDP growth one, two, and four steps ahead. The \textbf{left panels} present the \textit{cumulative} sum of forecast contributions $c_{ji}$ over the training sample (1961Q2 to 2019Q4), which collectively sum to the final predicted value $\hat{y}_j$ shown as dots. We initialize $c_{j0}$ at the unconditional average of the sample and present $c_{ji}$ as deviations from this average. The holdout sample ranges from 2020Q1 to 2025Q1, indicated by the dashed line. The \textbf{right panels} show forecast weights $w_j$ as a moving average of four quarters.  Lavender shading corresponds to NBER recessions.}
    \end{tablenotes}
  \end{minipage}
  \end{threeparttable}
\end{figure}

\begin{figure}[H]
\caption{\normalsize{Dual Interpretation of Post-Pandemic Recession Probabilities ($h=3$)}} \label{fig:classrec3}
\centering
\vspace*{-0.5em}
      \includegraphics[width=\textwidth, trim = 0mm 0mm 3mm 0mm, clip]{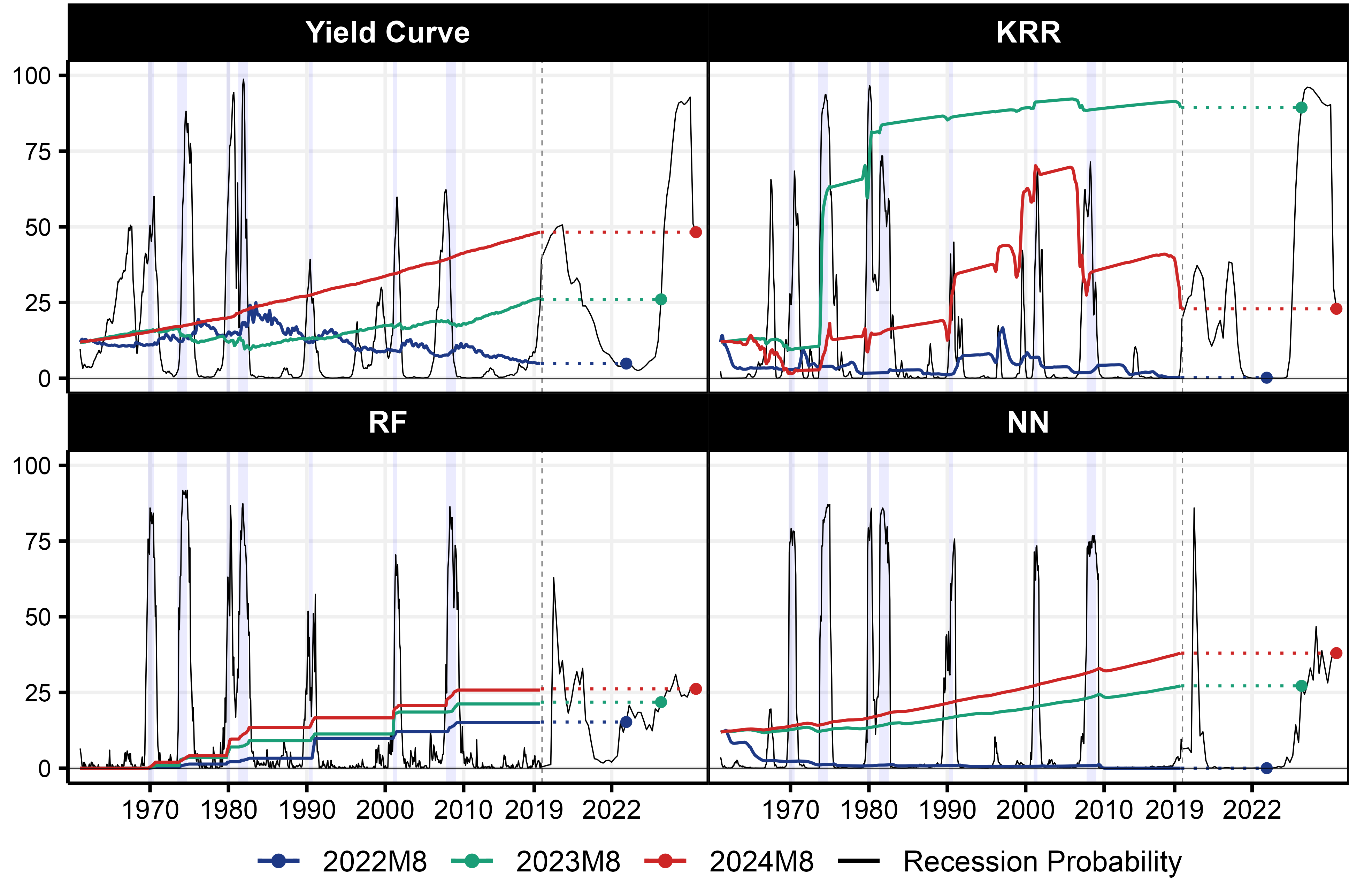}

 \begin{threeparttable}
    \centering
    \vspace*{-1.5em}
    \begin{minipage}{\textwidth}
      \begin{tablenotes}[para,flushleft]
    \setlength{\lineskip}{0.2ex}
    \footscript 
  {\textit{Notes}: The figure presents results from predicting recession probabilities three steps ahead. We present the \textit{cumulative} sum of forecast contributions $c_{ji}$ over the training sample (1961M4 to 2019M12), which collectively sum to the final predicted value $\hat{y}_j$ shown as dots. We initialize $c_{j0}$ at the unconditional average of the sample and present $c_{ji}$ as deviations from this average. Up to 2020M1 we plot in-sample results and show out-of-sample predictions through to 2024M8, indicated by the dashed line. Lavender shading corresponds to NBER recessions, i.e., the target variable.}
    \end{tablenotes}
  \end{minipage}
  \end{threeparttable}
\end{figure}

\end{document}